# Optimization of graded filleted lattice structures subject to yield and buckling constraints

Xiaoyang Wang[a,b], Lei Zhu[a], Liao Sun[b], Nan Li[a,*]

[a] *Dyson School of Design Engineering, Imperial College London, London, SW7 2AZ, UK*
[b] *The First Aircraft Institute of AVIC, Xi'an, 710087, China*

**Abstract.** To reduce the stress concentration and ensure the structural safety for lattice structure designs, in this paper, a new optimization framework is developed for the optimal design of graded lattice structures, innovatively integrating fillet designs as well as yield and elastic buckling constraints. Body-centred-cubic (BCC) lattices and primitive-cubic (PC) lattices are adopted in this study. Fillets are designed at the joints of the lattice struts. Both strut and fillet radii are defined as design variables. Homogenization method is employed to characterize the effective mechanical properties (i.e. elastic constants and yield stresses) of the lattice metamaterials, assisted by finite element analysis (FEA). Metamaterial models are developed using regression to represent the relationships between the metamaterial effective properties and the lattice geometric design variables. A yield constraint, based on the modified Hill's yield criterion, is developed and determined as a function of relative strut radii and fillet parameters, for the optimization of both the BCC and PC lattice structures. An elastic buckling constraint, based on the Euler buckling formula and the Johnson formula, is developed and determined as a function of relative strut radii, for the optimization of the PC lattice struts. Both yield and buckling constraints are integrated into an optimization problem formulation and a new optimization framework is proposed. The proposed optimization framework is tested by a case study of a compliance minimisation problem of a Messerschmitt-Bolkow-Blohm (MBB) beam. The result shows that the yield and buckling constraints guarantee the safety of the optimized structures, by ensuring no modified Hill's stresses or buckling factors at any lattice cell go beyond the yield and elastic buckling criteria, respectively. Compared with optimized graded lattice structures without fillets, 7% and 6% structural compliance reductions are achieved for the optimized beam structures composed of filleted BCC lattices and filleted PC lattices, respectively; in addition, reduced stress concentration regions are observed for the filleted lattice structures.

---

[*] Corresponding author
 E-mail address: n.li09@imperial.ac.uk *(Nan Li)*

## 1. Introduction

Inspired by natural hierarchical lightweight structures, such as bones and bamboo, periodic truss-like lattice structures have been widely used in the aerospace, automotive, and medical industries for improving structural efficiency [1]. These lattice structures consist of interconnected struts which are periodically arranged in a three-dimensional space. The lattice structures demonstrate various desirable physical properties, such as high stiffness-weight ratio, high energy-absorption capacity, outstanding thermal management capabilities, and excellent noise reduction capability, etc [2]. In addition, lattice structures can be optimized according to local load conditions by tailoring the distribution of their geometric parameters[3]. The recent development of additive manufacturing (AM) technologies, such as Selective Laser Melting (SLM) or Electron Beam Melting (EBM), enables the fabrication of lattice structures with complicated and novel architectures [4, 5]. Despite the promising properties of lattice structures, stress concentration at the strut joints of conventional lattice structures can lead to yield, creep, damage, or premature failure, and can cause a significant decrease in their mechanical performance [6-8]. Strut buckling caused by excessive axial load on thin struts is also a common failure mode of lattice structures [9]. Yield and buckling, as two typical types of failure modes, should be considered in the structural design and optimization process to ensure structural safety. In order to take advantage of the complicated geometry manufacturing capability of AM, a new methodology for filleted lattice structural optimization with yield and buckling constraints is proposed to generate graded lattices.

Since the pioneering work by Bendsøe and Kikuch in 1988 [10], Topology Optimization (TO) methods have experienced extensive development, particularly in recent years, because of the rapid development of AM technologies. Some dedicated reviews can be found in (van Dijk et al. [11], Liu et al [12]). However, topology optimized structures are too organic, so that a great number of supports are needed to ensure that overhang areas can be successfully built during the AM process. Printing and removing these support structures result in the waste of a lot of manufacturing costs and time. Since lattice structures can avoid requiring additional support structures in the AM process, design and optimization for lattice structures can potentially have more practical engineering applications. For example, a functionally graded lattice structures (FGLS) optimization method was proposed to infill a component with graded lattices, which could achieve multifunctional structures with the desired material distribution whilst being support-free for AM [13-17]. A lattice cell can be regarded as a type of truss-like artificial lattice metamaterial when the length of the lattice cell is much lower than the length of the component [18]. By optimizing metamaterials, the distribution of local microstructures can be tailored to achieve the optimal performance of the macroscale structure. Cheng et al. and Li et al. used an Asymptotic Homogenization (AH) method to build a lattice metamaterial for the optimization of lattice structures [16, 18]. Wang et al. proposed a numerical homogenization method and a gradient-free scheme of topology optimization to optimize the relative density distribution of the lattice metamaterial, to reduce the stress shielding of a hip implant [19]. However, in these designed and optimized lattice structures, the intersections of struts form sharp joints, which could lead to high stress concentration around the joints. Most studies on optimization of lattice structures have not taken the geometry of joints as design variables. Furthermore, most of them have focused on only stiffness optimization without considering the yield and buckling constraints.

A number of studies have been conducted to investigate the effect of lattice joints on the mechanical performance of lattice structures. S. P et al. proposed a new homogenization approach to estimate the mechanical properties of lattice structures whilst simultaneously considering joint stiffening effects. The results showed that a bending dominated lattice structure is more sensitive to joint properties than a stretching dominated lattice structure [6]. To reduce stress concentration, the designer should eliminate sharp geometry changes, particularly at lattice joints. L. Bai et al. proposed a new graded-strut body-centred cubic (GBCC) lattice structure with increased corner radii at the BCC nodes, which could increase the initial stiffness and plastic failure strength by at least 38.2% and 34.12% respectively when compared to the conventional BCC design [7]. X.F. Cao et al. introduced variable cross sections of struts to generate a new modified rhombic dodecahedron (RD) lattice structure, and both simulations and experiments showed that the modified RD lattice exhibited better mechanical properties and energy absorption compared with the conventional RD lattice structure [8]. Other related studies also showed that, by changing the geometry of the lattice joints to make the structural transition smoother, the mechanical properties of the lattice cell can be effectively improved [20-22]. However, by introducing new geometric parameters of lattice joints to lattices, the additional design variables complicate the optimal design problem. The design of filleted lattice also brings new challenges in lattice metamaterial modelling and CAD model generation, before and after optimization, respectively.

Lattice structures can fail either by plastic yielding or elastic buckling in engineering applications [9, 18]. Labeas and Sunaric investigated the failure responses of three different lattice types and developed a methodology comprising linear static and eigenvalue buckling analysis [23]. Souza et al. used an analytical method to establish yield criteria of an octet-truss lattice metamaterial, based on the Hill's yield criterion [9]. Deshpande et al. used an analytical method to develop an elastic buckling criterion for a f2BCCZ lattice metamaterial, based on the Euler buckling criterion [24]. Cheng et al. proposed a numerical method to establish an anisotropic yield criterion of a lattice structure and applied the yield constraints into the optimization of the lattice structure [16]. B. Ji et al. introduced the critical loads corresponding to Euler buckling and shear buckling as constraints to the optimization of a lattice sandwich plate [25]. However, few studies considered the yield and buckling constraints simultaneously in an optimization process.

The main objective of this paper is to propose methodologies and an optimization framework that enables the optimal design of graded lattice structures with filleted joints, and takes into account yield and buckling constraints simultaneously in an optimization process to maintain the safety of the optimized lattice structure. This paper is organized as follows: in Section 2, the selected optimization scheme using the periodic RVE homogenization method is described; in Section 3, the filleted BCC and PC lattice design methods are presented, including the fillet parameter definition and relative density calculation; in Section 4, the methods of characterizing the effective elastic properties of the lattice metamaterials based on numerical homogenization and developing metamaterial models through regression are introduced; in addition, the effects of the fillet parameter on the effective elastic moduli of the lattice metamaterials are analyzed; Section 5 presents the developments of the modified Hill's yield constraint and the elastic buckling constraints for lattice structural optimization; in Section 6, the optimization problem for minimizing structural compliance of graded filleted lattice structures is formulated; Section 7 shows the finally achieved optimization framework; in Section 8, a case study on minimising the compliance of an MBB beam is conducted to demonstrate the capability of the proposed optimization framework and the advantage of incorporating fillet designs; this is followed by the conclusion in Section 9.

## 2. A brief on the adopted optimization scheme

As is mentioned in section 1, lattice structures are multiscale structures. The macroscale structures are composed of numerous periodic microstructures, i.e. lattice cells. However, directly modelling detailed multiscale lattice structures is exceedingly computationally inefficient due to the extremely fine discretization requirement for modelling the detailed structures of the lattice cells. Therefore, the periodic RVE homogenization method is employed in this work to approximate the effective mechanical properties of lattice metamaterials [26]. Metamaterial models are developed to quantify the relationships between effective mechanical properties and design variables. The metamaterial models are then used in the optimization to avoid heavy computational burden.

The optimization procedure for a graded filleted lattice structure is illustrated in Figure 1. A lattice structure with a volume of $\Omega_\varepsilon$ is subjected to a traction $f_t$ at the structural boundary of $\Gamma_t$, and a displacement constraint $d$ on the boundary $\Gamma_d$. A body force $f_v$ is uniformly distributed in the domain $\Omega_\varepsilon$. The lattice cell is determined to be the RVE as shown in the figure. In this work, identical lattice cell sizes and lattice types are used to ensure structural integrity and good manufacturability. The effective mechanical properties of the lattice metamaterial are controlled by the geometric parameters of the lattice cell, i.e. strut radius and fillet radius. The metamaterial models are developed to bridge the optimization of macroscale structures and microscale lattice structures. Consequently, the distributions of the geometric parameters of lattice cells can be obtained by directly optimizing the homogenized macroscale structures.

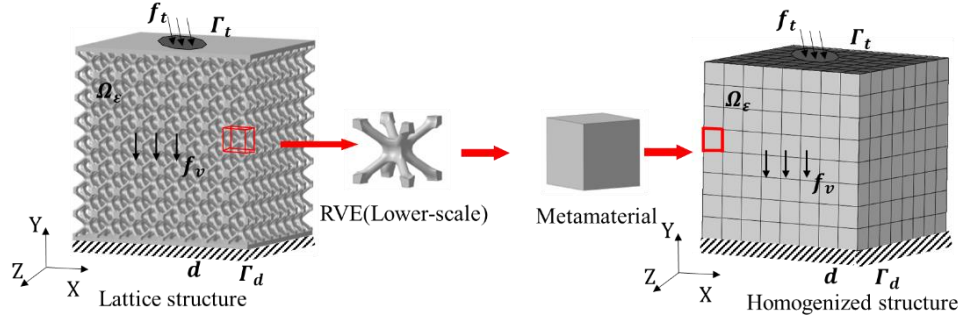

Figure 1.   Schematic illustration of the periodic RVE homogenization concept of a lattice structure

## 3. Design of BCC and PC lattice structures with fillets

The BCC and PC lattice types are investigated in this study. These two types of lattices are bending-dominant, according to Maxwell's criterion [27]. Therefore, severe stress concentration issues can occur at the joints of the lattice, which can consequently result in weakened effective mechanical properties [24, 28]. The sharp corners of the joints of these two types of lattice are filleted to reduce stress concentration and increase structural load bearing capability. The joint fillets of BCC and PC lattice cells are on their interior corners. Thus, the geometric function of a fillet is a concave function. The conventional and filleted BCC and PC lattice are illustrated in Figure 2. The figure demonstrates the modelling of filleted lattice cubes composed of 2×2×2 lattice cells. The size of the cubic lattice cell and the radius of the struts are $L$ and $R$ respectively. As shown in Figure 1. (a) and (c), the struts of the conventional lattice structures intersect each other, forming sharp geometry changes around the joints. Figure 2. (b) and (d) depict the schematics of smooth joints with fillets radii r.

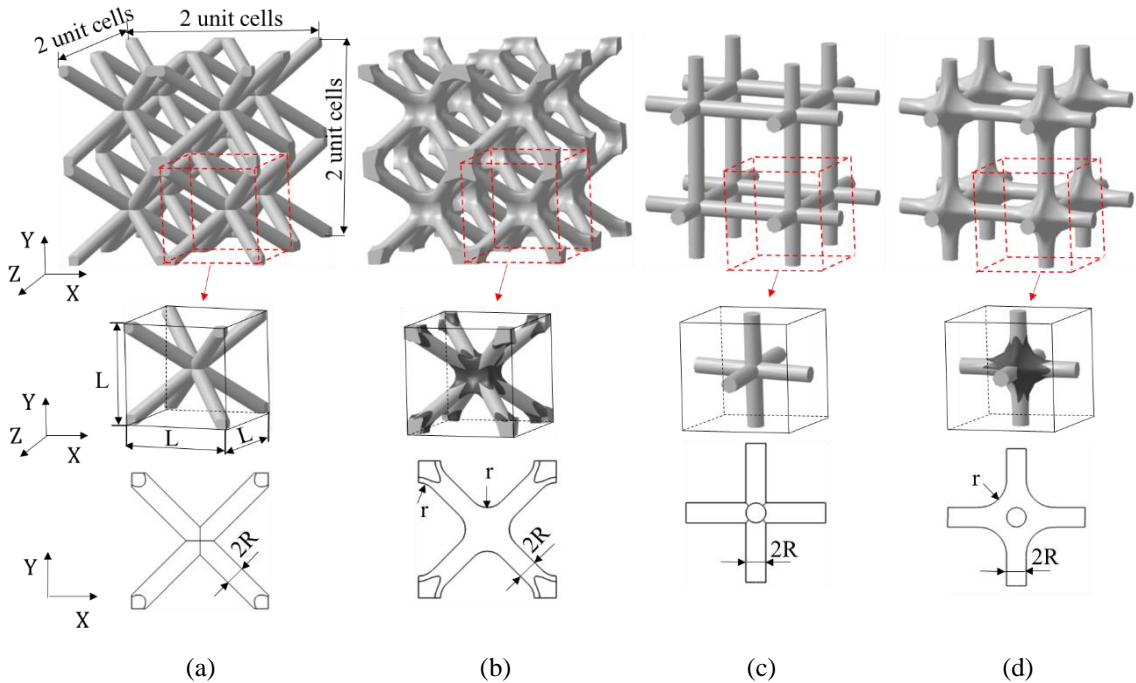

(a)                    (b)                    (c)                    (d)

Figure 2.   Conventional and filleted lattice structures. (a) Conventional BCC lattice structure with sharp joints (b) Filleted BCC lattice structure with smooth joints (c) Conventional PC lattice structure with sharp joints (d) Filleted PC lattice structure with smooth joints

The geometry of a certain type of lattice can be determined by its relative lattice cell density, $\bar{\rho}$, and relative strut radius, $\bar{R}$, which are defined by Eq. (1) and Eq. (2) respectively:

$$\bar{\rho} = \rho^H/\rho_s \tag{1}$$

$$\bar{R} = R/L \tag{2}$$

where $\rho^H$ is the effective density of the lattice cell and $\rho_s$ is the density of the parent material. The relative density can also be referred to as the volume fraction of a lattice cell. $R$ is the absolute radius of a lattice strut and $L$ is the length of the lattice cell.

The rolling ball blends method is a widely used geometric modelling method to generate fillets in sharp corners in Computer Aided Design (CAD) [29-31]. Many CAD software enable the parametric design of fillets. The fillets can be generated by defining their positions and radii. The radius of the fillet of a lattice cell is limited by the topological shape of the lattice cell and the void space in the lattice cell. A larger strut radius will leave a smaller space for fillet generation. It is observed from FE analysis that a lattice with a smaller strut radius is more sensitive to the stress concentration issue, hence requiring a larger fillet radius, i.e. the optimal fillet radius is dependent on the lattice strut radius. The effect of fillet radius on elastic moduli and yield stress will be discussed in detail in Sections 4 and 5, respectively. A fillet parameter $n$ is proposed in this study to represent the relationship between the fillet radius and the strut radius of a lattice cell. The relationships for BCC and PC lattices are defined in Eq. (3) and Eq. (4), respectively:

$$\bar{r}_{BCC} = \frac{r_{BCC}}{L} = \frac{n}{500\bar{R}} \quad (0 \leq n \leq 5) \tag{3}$$

$$\bar{r}_{PC} = \frac{r_{PC}}{L} = \frac{n(1-\bar{R})}{10} \quad (0 \leq n \leq 5) \tag{4}$$

Where $\bar{r}_{BCC}$ and $\bar{r}_{PC}$ are the relative fillet radii of the BCC and PC lattices, respectively; $r_{BCC}$ and $r_{PC}$ are the corresponding absolute fillet radii; $\bar{R}$ is the relative strut radius, and $n$ is the proposed fillet parameter. Thereby, the fillet radius is determined by both the fillet parameter and the lattice strut radius; it increases with increasing fillet parameter $n$ and decreasing strut radius. The range of $n$ in this study is between 0 and 5. When $n=0$, the lattice is the same as the conventional lattice without fillets.

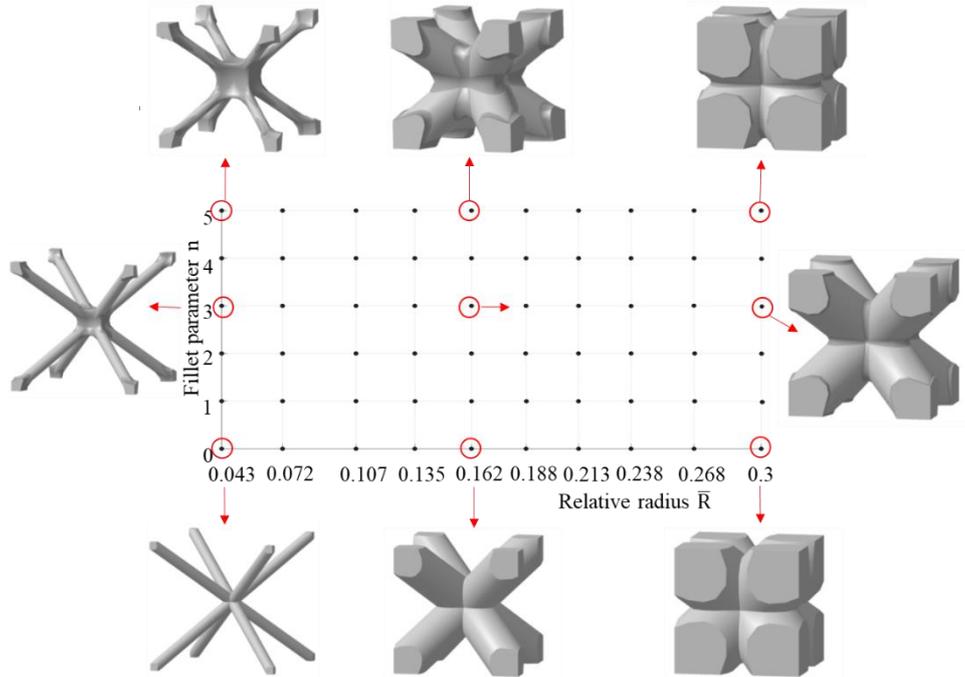

Figure 3. Sample points selection in the design domain of BCC lattices

The CAD models of the filleted lattice cells are built in CATIA V5R18. The volumes of the solid lattice CAD models can be directly extracted from the software and then the relative density of a lattice metamaterial, which is equal to the volume fraction of a lattice cell, can be calculated. Metamaterial models are built to quantify the relationship between the relative density and the geometric parameters ($\bar{R}$ and $n$) of the two types of lattice metamaterials. The full factorial sampling method is adopted to select the combinations of $\bar{R}$ and $n$. The distribution of samples is shown in Figure 3. The lower and upper bounds of relative density are defined, respectively, as 0.037 and 0.9 for both lattice types. This is based on considering the manufacturability of Additive Manufacturing. If the relative density is too low, the struts with excessively small radii and high aspect ratios can experience poor printing quality [32]. Using lattice metamaterials with high relative density could also cause manufacturing issues, such as the formation of internal cavities, which can affect the actual performance of the structures [32]. Correspondingly, the ranges of relative strut radii are [0.043, 0.3] for BCC lattices and [0.064, 0.457] for PC lattices. Fillet parameter $n$ ranges from 0 to 5 for both BCC and PC lattices. 60 samples are selected for each lattice.

The third order polynomial regression is adopted to quantify the relationship between the relative density $\bar{\rho}$ and the geometric variables $\bar{R}$ and $n$. Figure 4 shows response surfaces and the regression function of $\bar{\rho}(\bar{R}, n)$ of both lattice types, along with the coefficient of determination (R-squared values).

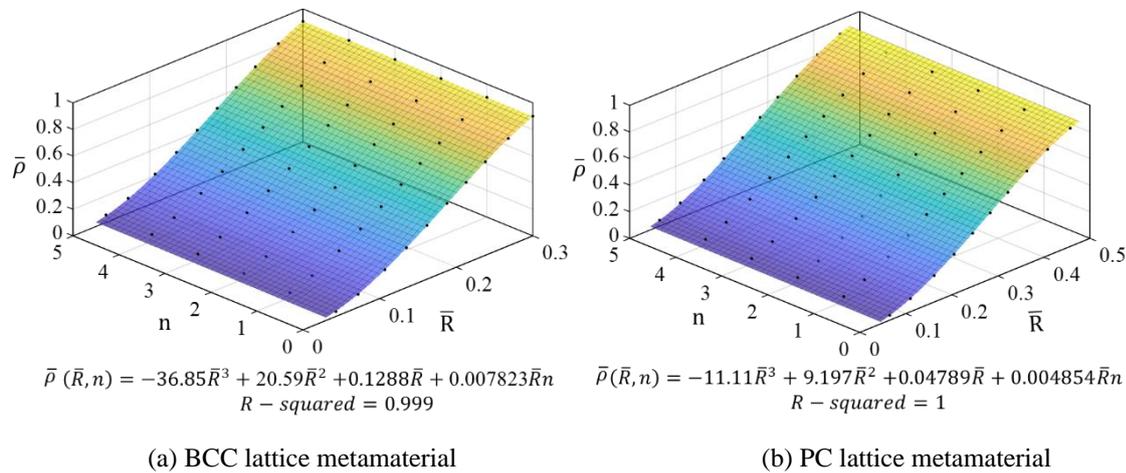

(a) BCC lattice metamaterial  (b) PC lattice metamaterial

Figure 4. Response surfaces of the relative density as a function of the strut relative radius and the fillet parameter for (a) BCC lattice; (b) PC lattice.

## 4. Lattice metamaterial models for effective elastic properties

### 4.1. Numerical homogenization

The effective mechanical properties of a lattice metamaterial are the main factors to be considered in solving lattice structure design and optimization problems. They are determined by the topological shape and the geometric parameters of a lattice cell. There are several approaches for calculating the effective mechanical properties of a lattice metamaterial, ranging from analytical and numerical methods to experimental investigations [18, 24, 33-35]. In analytical methods, the strut is regarded as a 1D rod for a stretching dominated lattice cell and a Euler–Bernoulli or Timoshenko beam for a bending dominated lattice cell to characterize the effective properties of the lattice cell [24, 33, 34]. It has been shown that these analytical methods are only suited for lattices with a high strut aspect ratio, and they ignore the lattice joint effects [33]. Among the numerical approaches, asymptotic homogenization (AH) theory and periodic representative volume element (RVE) homogenization have been widely applied to predict the effective mechanical properties of composite materials [36] and materials with periodic microstructures [37], and to solve multi-scale optimization problems [16, 18, 26]. The RVE is defined as the smallest representative material volume, and the macroscopic structure can be conceptualised as the stacking of these RVEs [36]. Additionally, these numerical methods have no limitation on the lattice cell shape and relative density.

For a macrostructure consisting of periodic arrays of repeated lattice cells, periodic boundary conditions (PBCs) should be prescribed to the RVE in the numerical homogenization method. Detailed discussions on the theory of PBC and homogenization is out of the scope of this paper. Readers can refer to Suquet [38]. For a cubic RVE, by applying displacement constraints equations to boundary nodes on opposite parallel surfaces of the RVE, PBCs can be easily applied in Finite Element Analysis (FEA) [26, 39]. This method ensures that there is no separation or overlap between the neighbouring lattice cells and can avoid over-constrained boundary conditions.

The numerical homogenization process is carried out in the FE solver of Abaqus 2018. After applying the PBCs and conducting prescript strain tests, the FEA results are post-processed to calculate the effective Young's modulus, effective shear modulus, and Poisson's ratio [26]. Figure 5 illustrates the undeformed model and deformations of a cubic lattice RVE with an initial size of $L_0$ subjected to a uniaxial tensile strain along axis-1 and a pure shearing strain in plane 1-2, respectively. $L_1$, $L_2$ and $L_3$ are the deformed dimensions of the RVE when it is under the uniaxial tensile strain. $H$ represents the maximum displacement value in direction-2 under pure shearing strain. In this study, the deformations of the lattice cells are restricted to small deformations in the linear elastic deformation domain. The effective Young's modulus in direction-1, $E_{11}^H$, is calculated by Eq. (5), where $\varepsilon_{11}^H$ is the imposed tensile strain of the RVE in direction-1 and $\sigma_{11}^H$ is the averaged stress calculated from the reaction force $F_1$ measured in the direction-1. The effective Poisson's ratio $\mu_{12}^H$ is the negative of the ratio of transverse strain $\varepsilon_{22}^H$ to longitudinal strain $\varepsilon_{11}^H$, as illustrated in Eq. (6). Similarly, the effective Poisson's ratio $\mu_{13}^H$ is calculated in Eq. (7). The effective shear modulus, $G_{12}^H$, is calculated in Eq. (8), where $\varepsilon_{12}^H$ is the imposed shear strain in plane 1-2 and $\sigma_{12}^H$ is the averaged stress calculated from the measured reaction force $F_{12}$.

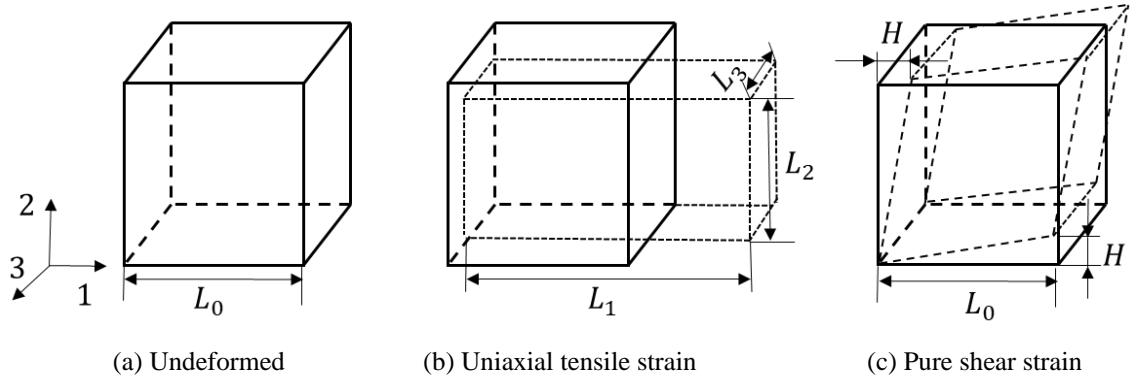

(a) Undeformed    (b) Uniaxial tensile strain    (c) Pure shear strain

Figure 5.   Demonstration of loading conditions for the periodic RVE homogenization method.

$$E_{11}^H = \frac{\sigma_{11}^H}{\varepsilon_{11}^H} = \frac{F_1/L_0^2}{(L_1-L_0)/L_0} \tag{5}$$

$$\mu_{12}^H = \frac{-\varepsilon_{22}^H}{\varepsilon_{11}^H} = \frac{L_0-L_2}{L_1-L_0} \tag{6}$$

$$\mu_{13}^H = \frac{-\varepsilon_{33}^H}{\varepsilon_{11}^H} = \frac{L_0-L_3}{L_1-L_0} \tag{7}$$

$$G_{12}^H = \frac{\sigma_{12}^H}{2\varepsilon_{12}^H} = \frac{-F_{12}/L_0^2}{2(H/L_0)} \tag{8}$$

For both the BCC and PC lattices defined in this study, since the struts have the same radius and the fillet parameters have the same value in any lattice cell, all lattice cells have three orthogonal symmetric planes. This type of RVE is equivalent to a quasi-isotropic metamaterial. Thus, its effective Young's moduli, effective Poisson's ratios, and effective shear moduli satisfy Eq. (9-11), respectively.

$$E_{11}^H = E_{22}^H = E_{33}^H = E^H \tag{9}$$

$$G_{12}^H = G_{23}^H = G_{13}^H = G^H \tag{10}$$

$$\mu_{12}^H = \mu_{13}^H = \mu_{21}^H = \mu_{23}^H = \mu_{31}^H = \mu_{32}^H = \mu^H \tag{11}$$

Therefore, the effective elasticity tensor of a quasi-isotropic lattice cell can be expressed using three unique elastic constants $C_{11}^H$, $C_{12}^H$, and $C_{44}^H$. The elastic constitutive equation is presented as Eq. (12), where the elastic constants are calculated by using Eq. (13).

$$\begin{bmatrix} \sigma_{11}^H \\ \sigma_{22}^H \\ \sigma_{33}^H \\ \sigma_{12}^H \\ \sigma_{23}^H \\ \sigma_{13}^H \end{bmatrix} = \begin{bmatrix} C_{11}^H & C_{12}^H & C_{12}^H & 0 & 0 & 0 \\ C_{12}^H & C_{11}^H & C_{12}^H & 0 & 0 & 0 \\ C_{12}^H & C_{12}^H & C_{11}^H & 0 & 0 & 0 \\ 0 & 0 & 0 & C_{44}^H & 0 & 0 \\ 0 & 0 & 0 & 0 & C_{44}^H & 0 \\ 0 & 0 & 0 & 0 & 0 & C_{44}^H \end{bmatrix} \begin{bmatrix} \varepsilon_{11}^H \\ \varepsilon_{22}^H \\ \varepsilon_{33}^H \\ 2\varepsilon_{12}^H \\ 2\varepsilon_{23}^H \\ 2\varepsilon_{13}^H \end{bmatrix} \tag{12}$$

$$\begin{cases} C_{11}^H = \dfrac{1-(\mu^H)^2}{(E^H)^2 \Delta} \\ C_{12}^H = \dfrac{\mu^H + (\mu^H)^2}{(E^H)^2 \Delta} \\ C_{44}^H = G^H \\ \Delta = \dfrac{1 - 3(\mu^H)^2 - 2(\mu^H)^3}{(E^H)^3} \end{cases} \tag{13}$$

*4.2. FEA mesh sensitivity analysis*

The FEA method is employed using the commercial software ABAQUS to calculate the effective mechanical properties of the RVE. Since numerical results are associated with mesh type and mesh size, mesh sensitivity analysis should be evaluated before conducting the numerical homogenization. In this study, three types of mesh, C3D4, C3D10, and C8D12, are applied in BCC and PC lattice RVE to study the mesh sensitivity on an FEA simulation to obtain a suitable mesh type and mesh size to balance accuracy and efficiency. C3D4 and C3D10 are tetrahedral meshes, and C8D12 is hexahedral mesh. Voxel modelling is used to generate the FEA model with C8D12 to adapt to the complex geometry of the lattice RVE [35]. $N_m$ represents the number of meshes along the radius of the strut cross section. $N_m$ is calculated by Eq. (14), where $R$ is the radius of the lattice strut and $M_{size}$ is the applied mesh size. Figure 6 illustrates the convergence history of the mesh sensitivity analysis of a BCC and PC lattice RVE, respectively. The normalized effective elastic constants $\bar{C}_{ij}$ can be expressed as Eq. (15), where $C_{ij}^H$ and $C_{ij}^*$ are the effective elastic constant of the RVE and the corresponding elastic constant of the parent material, respectively. It can be seen that when $N_m$ is 8, $\bar{C}_{ij}$ converged based on the tolerance of 5% for all three mesh types. Since the C3D4 mesh needs less nodes and integration points in the FEA simulation but achieves almost the same level of accuracy compared with the other two mesh types, in this study, C3D4 with $N_m = 8$ is used in the FEA simulation.

$$N_m = \frac{R}{M_{size}} \tag{14}$$

$$\bar{C}_{ij} = C_{ij}^H / C_{ij}^* \tag{15}$$

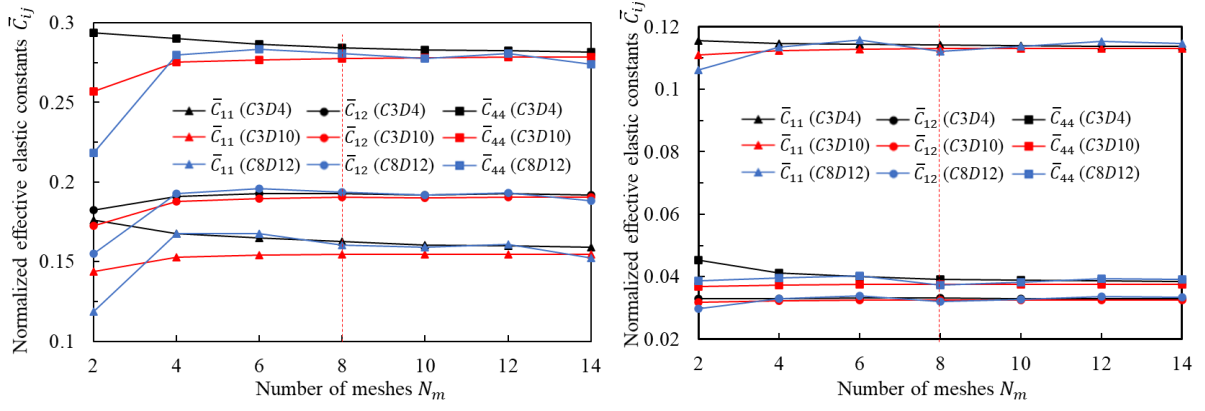

(a) BCC lattice RVE ($\bar{R}=0.213$, $n=0$)  (b) PC lattice RVE ($\bar{R}=0.245$, $n=0$)

Figure 6.  Mesh sensitivity analysis of the lattice RVE

*4.3. Lattice metamaterial models for effective elastic properties*

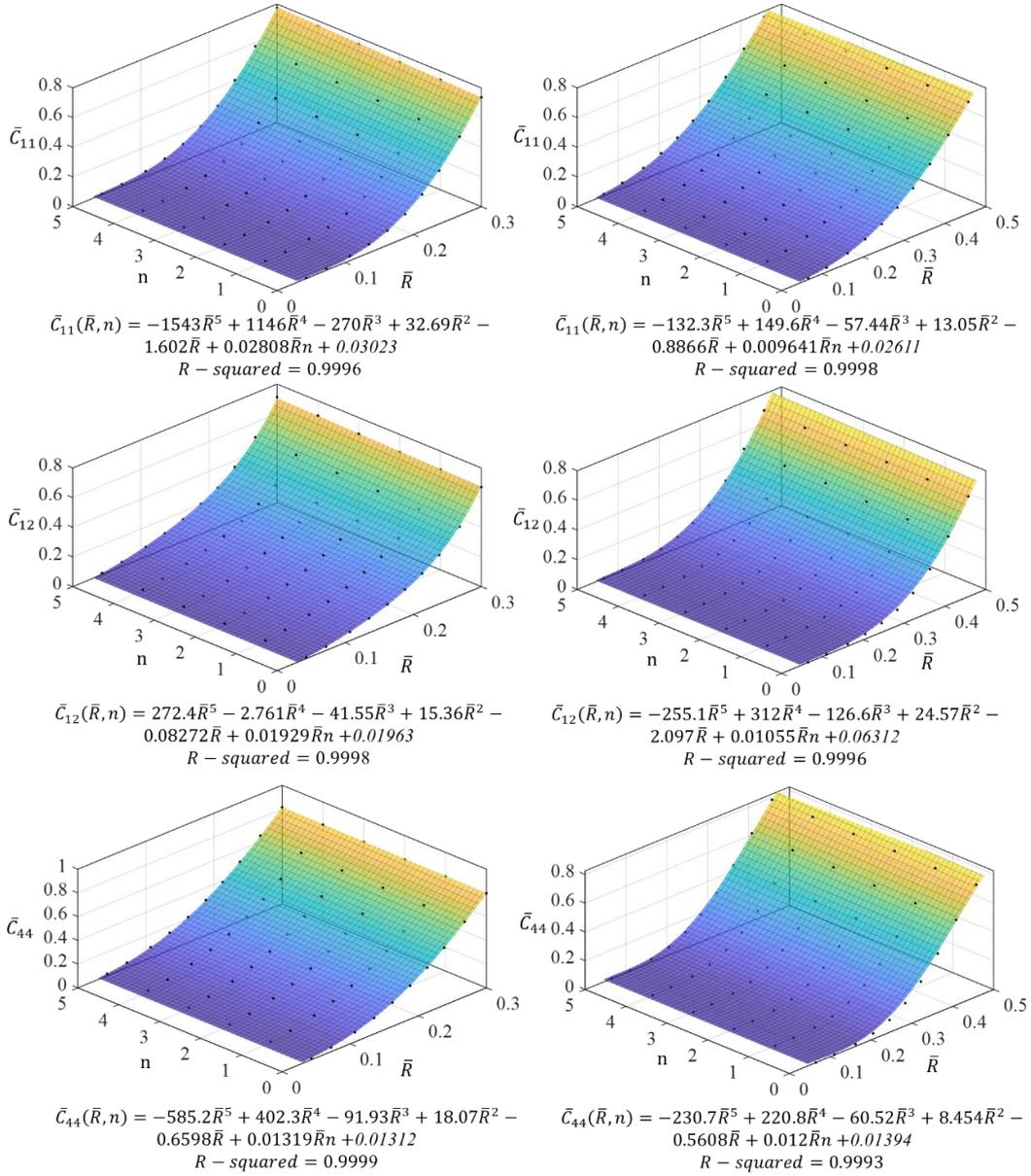

$\bar{C}_{11}(\bar{R},n) = -1543\bar{R}^5 + 1146\bar{R}^4 - 270\bar{R}^3 + 32.69\bar{R}^2 - 1.602\bar{R} + 0.02808\bar{R}n + 0.03023$
$R - squared = 0.9996$

$\bar{C}_{11}(\bar{R},n) = -132.3\bar{R}^5 + 149.6\bar{R}^4 - 57.44\bar{R}^3 + 13.05\bar{R}^2 - 0.8866\bar{R} + 0.009641\bar{R}n + 0.02611$
$R - squared = 0.9998$

$\bar{C}_{12}(\bar{R},n) = 272.4\bar{R}^5 - 2.761\bar{R}^4 - 41.55\bar{R}^3 + 15.36\bar{R}^2 - 0.08272\bar{R} + 0.01929\bar{R}n + 0.01963$
$R - squared = 0.9998$

$\bar{C}_{12}(\bar{R},n) = -255.1\bar{R}^5 + 312\bar{R}^4 - 126.6\bar{R}^3 + 24.57\bar{R}^2 - 2.097\bar{R} + 0.01055\bar{R}n + 0.06312$
$R - squared = 0.9996$

$\bar{C}_{44}(\bar{R},n) = -585.2\bar{R}^5 + 402.3\bar{R}^4 - 91.93\bar{R}^3 + 18.07\bar{R}^2 - 0.6598\bar{R} + 0.01319\bar{R}n + 0.01312$
$R - squared = 0.9999$

$\bar{C}_{44}(\bar{R},n) = -230.7\bar{R}^5 + 220.8\bar{R}^4 - 60.52\bar{R}^3 + 8.454\bar{R}^2 - 0.5608\bar{R} + 0.012\bar{R}n + 0.01394$
$R - squared = 0.9993$

(a) BCC lattice  (b) PC lattice

Figure 7.  Response surfaces of the normalized elastic constants, based on the determined metamaterial models of BCC and PC lattices.

To quantify the relationships between the normalized effective elastic constants of lattice metamaterials and the geometric parameters ($\bar{R}$ and $n$) of lattice cells, numerical homogenization is conducted on the sampled lattice cells for both PC and BCC metamaterials and regression models are developed. The polynomial regression function is provided in Eq. (16), where $a_1$ to $a_7$ are model coefficients to be determined for each normalized effective elastic constant $\bar{C}_{ij}$ of each metamaterial type. Figure 7 illustrates the response surfaces of normalized effective elastic constants of the BCC and PC lattice metamaterials. It can be observed that all response surfaces are smooth and R-squared values are all greater than 0.999, indicating high regression accuracy.

$$\bar{C}_{ij}(\bar{R}, n) = a_1 \bar{R}^5 + a_2 \bar{R}^4 + a_3 \bar{R}^3 + a_4 \bar{R}^2 + a_5 \bar{R} + a_6 \bar{R} n + a_7 \tag{16}$$

*4.4. Effects of the fillet parameter on effective elastic moduli of the lattice metamaterials*

The effects of the fillet parameter on effective Young's modulus $E^H$ and shear modulus $G^H$, at different relative density levels for each lattice type, are illustrated in Figure 8. For a clearer comparison, the ratios of the effective moduli of filleted lattices to the effective moduli of the un-filleted lattice (denoted as $E^H/E_0^H$ and $G^H/G_0^H$), for each relative density level of each lattice type, are used as the unified vertical axes in Figure 8. Figure 8 (a) shows the effects of the fillet parameter $n$ on the BCC lattice metamaterials. It can be observed that the effective Young's modulus of the BCC lattice metamaterial is more sensitive to the change in $n$, compared with the effective shear modulus $G^H$. For example, at the relative density of 0.2, when $n$ increases from 0 to 5, $E^H/E_0^H$ is improved by over 55%, while $G^H/G_0^H$ is only improved by 3.9%. However, unlike BCC, Figure 8 (b) shows that, for the PC lattice metamaterials, the effective shear modulus is more sensitive to the change in $n$ compared with the effective Young's modulus $E^H$. For example, at the relative density of 0.2, when $n$ increases from 0 to 5, $G^H/G_0^H$ increases by over 40%, while $E^H/E_0^H$ remains at almost the same level. For both BCC and PC, the improvements in the effective moduli resulting from the increase of the fillet parameter $n$ become less obvious with increasing relative density of the lattice metamaterials.

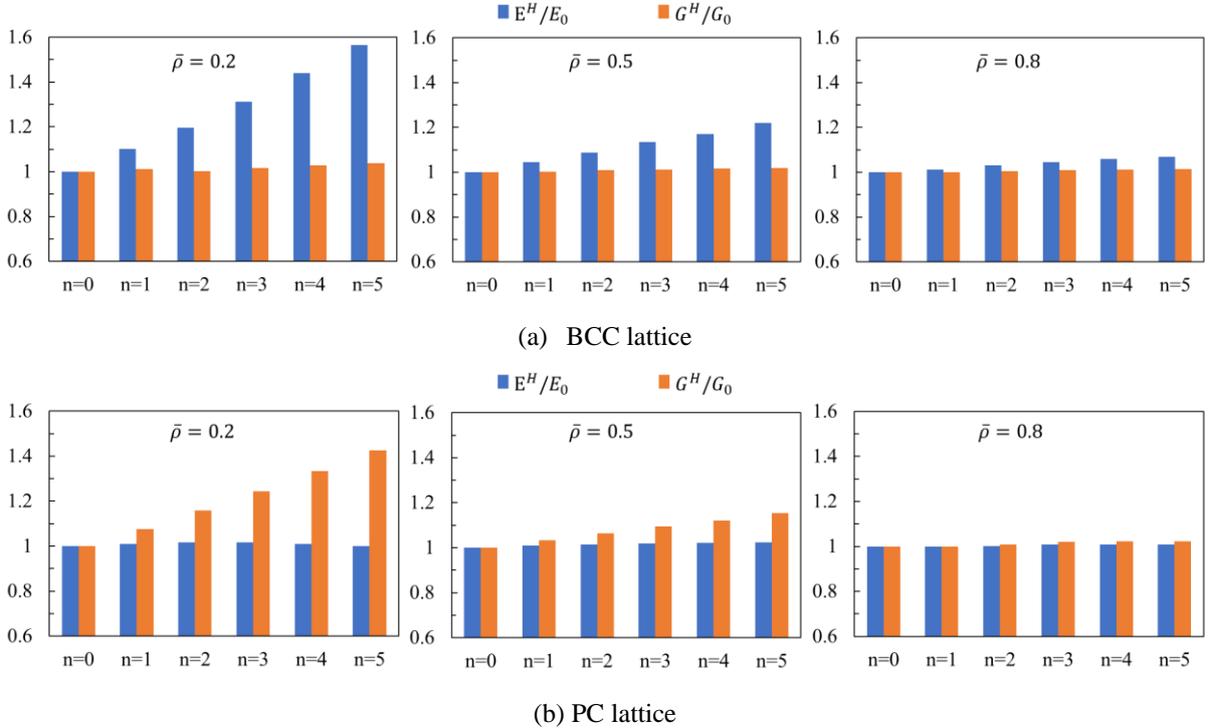

Figure 8. The effects of the fillet parameter on the effective elastic moduli of the lattice metamaterials

*4.5. Lattice cell size sensitivity analysis*

As mentioned in Section 2, the homogenization method assumes that the lattice cell is sufficiently small in size compared to the macroscale structure. However, due to the printing resolution of AM, the lattice cell cannot be arbitrarily small. For better manufacturing efficiency and quality, a bigger size of lattice cells is desirable. Therefore, the effects of lattice cell size on the accuracy of numerical homogenization are investigated through FEA simulations, which can help to determine the cell size for lattice structure design optimization and manufacturing. Figure 9 (a) shows the loading conditions, uniaxial compression, and simple shear of a cubic structure. Here, a BCC lattice cell with relative radius $\bar{R} = 0.213$ and fillet parameter $n = 0$ is used to compose the cubic structure. By varying the cell size, a range of detailed FEA models of lattice structures are generated. $N_{cell}$ denotes the number of lattice cell along one direction of the cubic structure. Simulations of each detailed model under both loading conditions are conducted, to compare with the results of the homogenized model where the corresponding metamaterial model is applied to the cubic structure. The strain energy percentage difference between the homogenized model and the detailed model is calculated through equation (17) and is used as the criteria to determine the minimum $N_{cell}$.

The strain energy difference is defined as:

$$\Delta \text{SE} = |SE_H - SE_d|/SE_d \cdot 100\% \tag{17}$$

where $SE_H$ and $SE_d$ denote the strain energy of the homogenized model and the detailed model, respectively.

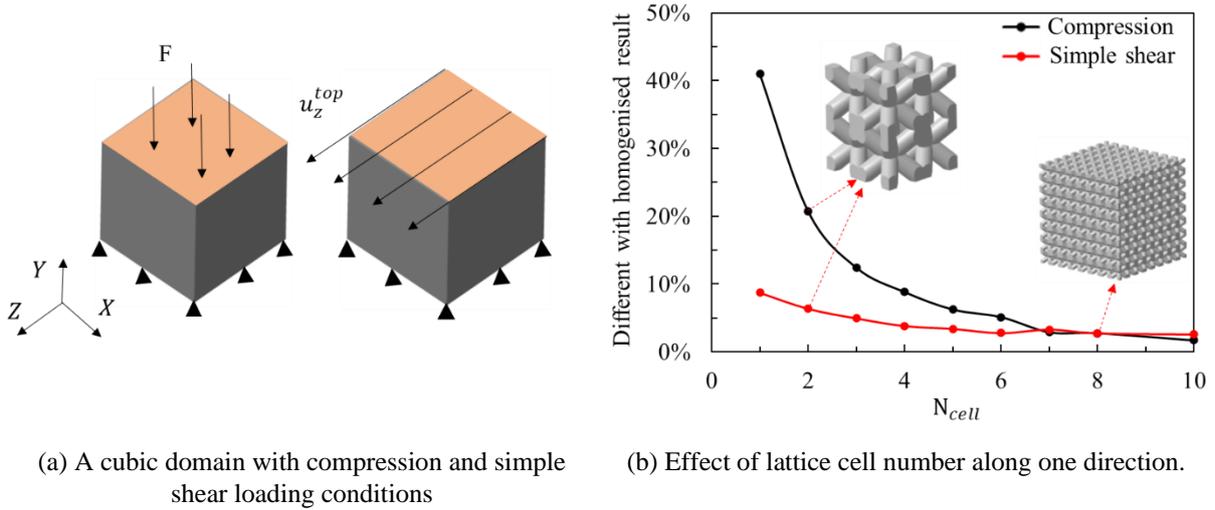

(a) A cubic domain with compression and simple shear loading conditions

(b) Effect of lattice cell number along one direction.

Figure 9.    Characterizing of effect of the lattice cell size.

Figure 9 (b) presents the effect of lattice cell number (inversely related to lattice cell size) on the strain energy percentage difference between the homogenized model and the detailed model. It can be observed that ΔSE converges when $N_{cell}$ reaches 8, and the converged value of ΔSE is within 5%. It indicates that the accuracy of the homogenization method satisfies a general engineering requirement when $N_{cell}$ reaches 8 and above. Then, $N_{cell} = 8$ is used to test PC lattices with relative radius $\bar{R} = 0.283$ and fillet parameter $n = 0$, and the values of ΔSE are 1% and 4.16% for compression test and simple shear test, respectively. Thus, in this study, the minimum number of lattice cells along one direction of the design component is determined to be 8.

**5. Development of yield and buckling criteria for lattice structural optimization**

*5.1. Modified Hill's yield criterion for lattice metamaterials*

A yield constraint for the proposed lattice structural optimization platform is developed in this Section to ensure the structural safety of the optimized lattice structures. The Hill's yield criterion is a widely used yield criterion for describing anisotropic plastic deformations. A modified Hill's yield criterion is proposed by Deshpande et al. to describe the yield behaviour of lattice metamaterials with orthotropic symmetry [24]. This modified Hill's yield criterion is employed in this study to formulate the yield constraint for optimization. Assuming the yield stresses of the lattice metamaterial under compression and tension are equal, because the principal axes of anisotropy coincide with the reference axes $(x_1, x_2, x_3)$ for the studied lattice types, the modified Hill's yield criterion [24] can be formulated as:

$$\emptyset \equiv (\sigma_d^H)^2 + P(\sigma_m^H)^2 - 1 = 0 \tag{18}$$

Where $\sigma_d^H$ and $\sigma_m^H$ are the deviatoric stress and hydrostatic pressure applied to the lattice metamaterials, respectively. P is a material constant. $\sigma_d^H$ and $\sigma_m^H$ can be calculated using Eq. (19) and Eq. (20), respectively.

$$(\sigma_d^H)^2 = A_1(\sigma_{11}^H - \sigma_{22}^H)^2 + A_2(\sigma_{22}^H - \sigma_{33}^H)^2 + A_3(\sigma_{11}^H - \sigma_{33}^H)^2 + B_1(\sigma_{12}^H)^2 + B_2(\sigma_{23}^H)^2 + B_3(\sigma_{13}^H)^2 \tag{19}$$

$$\sigma_m^H = \frac{\sigma_{11}^H + \sigma_{22}^H + \sigma_{33}^H}{3} \tag{20}$$

Where $A_1, A_2, A_3, B_1, B_2$ and $B_3$ are material constants and $\sigma_{ij}^H (i,j = 1,2,3)$ represents the effective stress of the lattice cell. Because the designed BCC and PC lattices in this study are quasi-isotropic metamaterials, the seven material constants can be reduced to three by applying the relations of $A_1 = A_2 = A_3 = A$ and $B_1 = B_2 = B_3 = B$. Hence, the modified Hill's yield criterion can be rewritten in a matrix form:

$$\emptyset(\boldsymbol{\sigma}^H) \equiv (\boldsymbol{\sigma}^H)^T \boldsymbol{M} \boldsymbol{\sigma}^H - 1 = (\sigma^{Hill})^2 - 1 \tag{21}$$

$$\boldsymbol{M} = \begin{bmatrix} 2A + P/9 & P/9 - A & P/9 - A & 0 & 0 & 0 \\ P/9 - H_1 & 2A + P/9 & P/9 - A & 0 & 0 & 0 \\ P/9 - A & P/9 - A & 2A + P/9 & 0 & 0 & 0 \\ 0 & 0 & 0 & B & 0 & 0 \\ 0 & 0 & 0 & 0 & B & 0 \\ 0 & 0 & 0 & 0 & 0 & B \end{bmatrix} \tag{22}$$

where $\sigma^{Hill}$ represents the modified Hill's stress. The lattice metamaterial yields when $\sigma^{Hill} \geq 1$. The three material constants A, B, and P, are functions of yield stresses of the lattice metamaterials defined as:

$$P = \frac{1}{(\sigma_m^{HY})^2} \tag{23}$$

$$A = \frac{1}{2(\sigma_{11}^{HY})^2} - \frac{P}{18} \tag{24}$$

$$B = \frac{1}{(\sigma_{12}^{HY})^2} \tag{25}$$

where $\sigma_{11}^{HY}$, $\sigma_{12}^{HY}$ and $\sigma_m^{HY}$ are the effective uniaxial yield stress, effective shear yield stress, and effective hydrostatic yield stress of the lattice metamaterials, respectively. The three effective yield stresses of lattice metamaterials can be obtained through conducting three types of numerical experiments on the lattice RVEs with the PBCs applied. The three types include the uniaxial tensile test in direction-1, the pure shear test in plane 1-2, and the hydrostatic compression test. In this paper, Al 2024-T3 is used as the parent material that the lattice metamaterial is composed of. The parent material is isotropic and is simplified as a bilinear elastic plastic material with a constant tangent modulus. For the parent material, the Young's modulus $E^*$ is 70 GPa, the Poisson's ratio $\mu^*$ is 0.3, yield stress $\sigma_y^*$ is 326 MPa, and tangent modulus $E_t^*$ is 4746.7 MPa. Albeit the standard offset method to define yield strength is at 0.2% offset, a more conservative value, i.e. a 0.02% offset, is defined for defining the yield strength of the lattice RVEs in this study due to safety concerns which may be caused by additive manufacturing defects. The commercial FE software Abaqus 2018 is used to conduct the aforementioned three types of numerical experiments on the previously defined sets of lattice RVE samples. The obtained results are used to model the three-independent effective yield stresses ($\sigma_{11}^{HY}$, $\sigma_{12}^{HY}$ and $\sigma_m^{HY}$) as the function of strut relative radius $\bar{R}$ and fillet parameter $n$. For each lattice type, fourth order polynomial regressions are employed to achieve the metamaterial models for the normalized effective yield stresses, as Eq. (26-28), where $\sigma_y^*$ is the yield stress of the parent material.

$$\bar{\sigma}_{ij}^Y(\bar{R}, n) = \frac{\sigma_{11}^{HY}}{\sigma_y^*} = \frac{\sigma_{22}^{HY}}{\sigma_y^*} = \frac{\sigma_{33}^{HY}}{\sigma_y^*} = b_1\bar{R}^4 + b_2\bar{R}^3 + b_3\bar{R}^2 + b_4\bar{R} + b_5\bar{R}n \quad (i = j\,;\, i,j = 1,2,3) \quad (26)$$

$$\bar{\sigma}_{ij}^Y(\bar{R}, n) = \frac{\sigma_{12}^{HY}}{\sigma_y^*} = \frac{\sigma_{23}^{HY}}{\sigma_y^*} = \frac{\sigma_{13}^{HY}}{\sigma_y^*} = c_1\bar{R}^4 + c_2\bar{R}^3 + c_3\bar{R}^2 + c_4\bar{R} + c_5\bar{R}n \quad (i \neq j\,;\, i,j = 1,2,3) \quad (27)$$

$$\bar{\sigma}_m^Y(\bar{R}, n) = \frac{\sigma_m^{HY}}{\sigma_y^*} = d_1\bar{R}^4 + d_2\bar{R}^3 + d_3\bar{R}^2 + d_4\bar{R} + d_5\bar{R}n \quad (28)$$

where $b_1$ to $b_5$, $c_1$ to $c_5$, $d_1$ to $d_5$, are model coefficients to be determined for each normalized effective yield stresses of each metamaterial type.

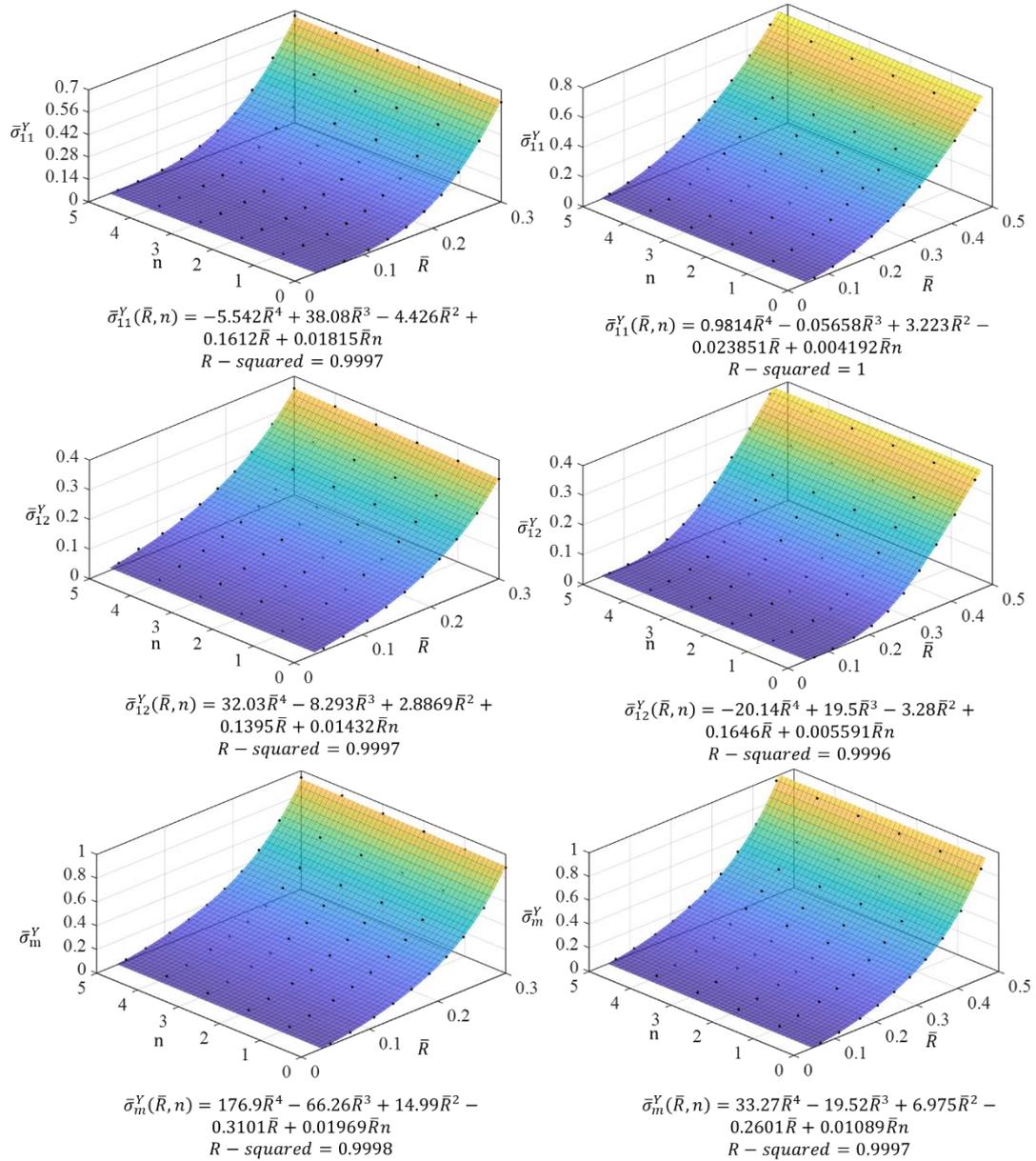

(a) BCC lattice  (b) PC lattice

Figure 10. Response surfaces of normalized effective yield stresses, based on the determined metamaterial models of BCC and PC lattices.

Figure 10 illustrates the response surfaces of the three normalized effective yield stresses based on the determined models of the BCC and PC lattice metamaterials. As can be seen, all response surfaces are smooth and the R-squared values for the regression results of the metamaterial models are all greater than 0.999, indicating a high accuracy of the regression.

## 5.2. Effects of fillet parameter on yield of lattice metamaterials

The effects of the fillet parameter $n$ on the effective yield stresses, at different relative density levels for each lattice type, are illustrated in Figure 11. For a clearer comparison, the ratios of the normalized effective yield stresses of filleted lattices to the normalized effective yield stresses of the un-filleted lattice (denoted as $\bar{\sigma}_{11}^Y/\bar{\sigma}_{11_0}^Y$, $\bar{\sigma}_{12}^Y/\bar{\sigma}_{12_0}^Y$, and $\bar{\sigma}_m^Y/\bar{\sigma}_{m_0}^Y$), for each relative density level of each lattice type, are used as the unified vertical axes in Figure 11. In general, the effects of increasing the fillet parameter $n$ on the effective yield stresses are more significant at a lower level of relative density, and become less obvious with the increase in the relative density of the lattice metamaterials. Figure 11 (a) shows the effects of $n$ on the yield stresses of the BCC lattice metamaterials. At the relative density level of 0.2, compared with the un-filleted lattice, the uniaxial yield stress and shear yield stress of the BCC lattice metamaterial are improved by more than 24% and 14%, respectively, when $n$ is increased 5, while the hydrostatic yield stress remains almost stable. Figure 11 (b) illustrates the effects of the fillet parameter $n$ on the yield stresses of the PC lattice metamaterials. At the relative density level of 0.2, the shear yield stress of the PC lattice metamaterial improves by over 25% when $n$ is increased to 5, compared with the un-filleted lattice; however, the uniaxial yield stress and hydrostatic yield stress moderately decrease with the increase of $n$.

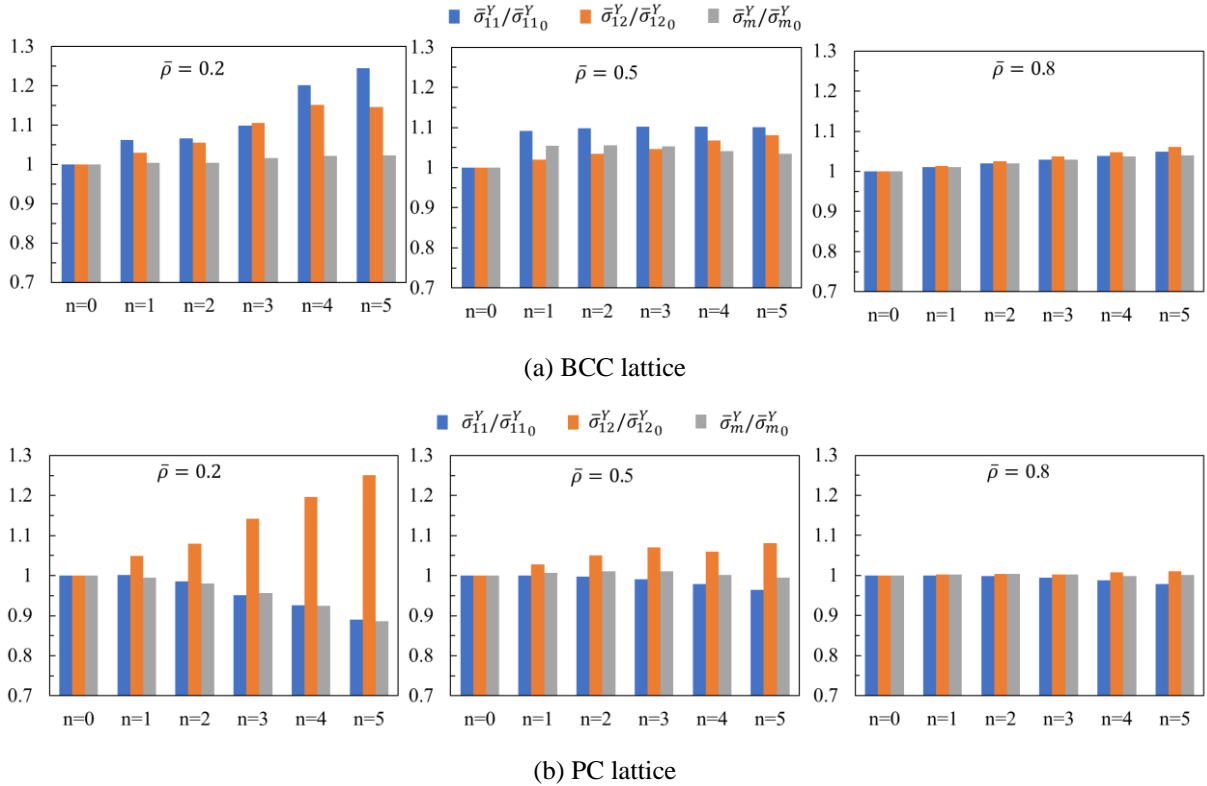

Figure 11. The effects of the fillet parameter on the normalized effective yield stresses of the lattice metamaterials.

The difference in the effects of the fillet parameter $n$ on the BCC and PC lattice metamaterials could be caused by the following two reasons. Firstly, it can be observed in Figure 12 that the struts of a BCC lattice cell mainly endure bending deflection under both uniaxial and shear loading conditions, causing a large stress concentration at the lattice joint. This is evidenced by the fact that when BCC lattice structures collapse, they typically show plastic deformation at the joints [7]. However, for the PC lattice metamaterial, a large stress concentration at the lattice joint occurs only when the lattice cell sustains a shear load. Because the maximum bending moment occurs at the root of a strut when it endures a bending deformation, high stress concentration is caused at the lattice joint. Accordingly, by increasing the fillet parameter $n$, the stress concentration can be significantly reduced, and consequently the uniaxial and shear yield stresses of the BCC lattice metamaterials and the shear yield stress of PC lattice metamaterials can be improved. Secondly, when the PC lattice is subjected to uniaxial or hydrostatic loads, the struts sustain axial force, so there is no stress concentration occurring at the joints of the lattice struts. To retain the same density level of PC lattice metamaterials, the increase in $n$ will cause a decrease in strut radius R; but R contributes more to the resistance of the axial load than $n$. Therefore, adding a fillet cannot improve the uniaxial or hydrostatic yield stress of the PC lattice metamaterials.

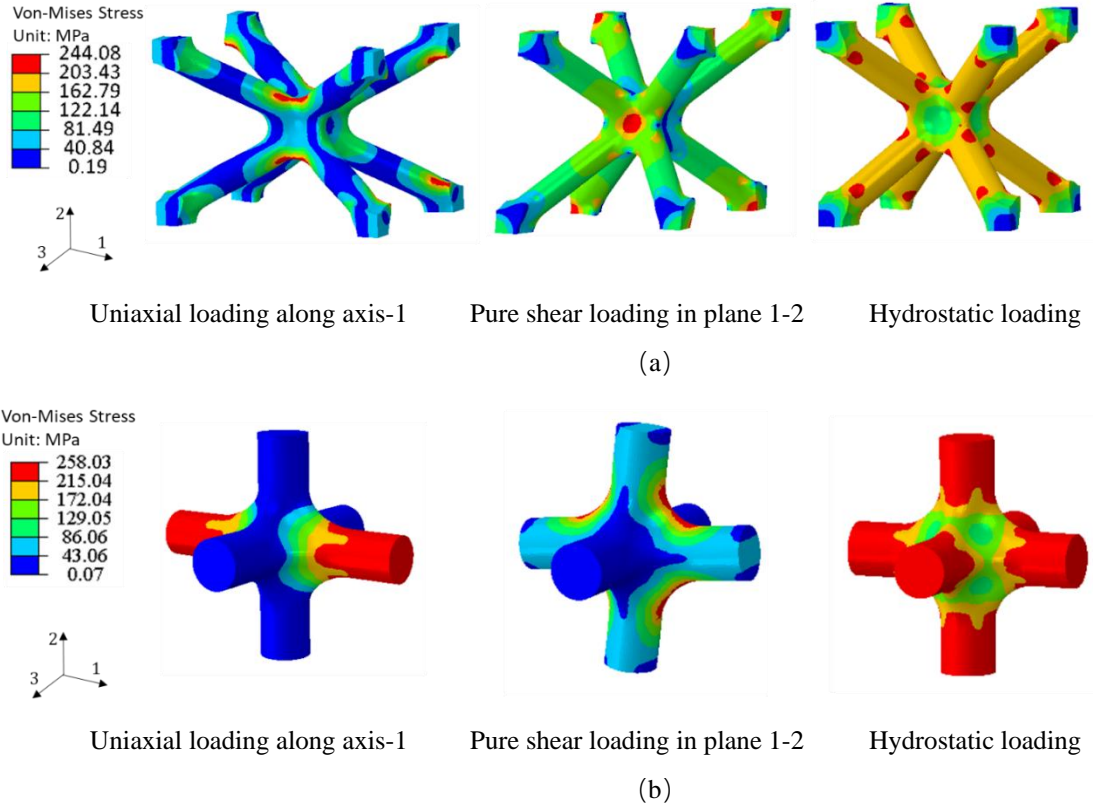

Figure 12. Von-Mises stress contours of example lattice RVEs: (a) BCC lattice ($\bar{R} = 0.072, n = 4$) and (b) PC lattice ($\bar{R} = 0.110, n = 4$), under three types of loading conditions in the numerical experiments.

*5.3. Development of buckling criteria for PC lattice struts*

As discussed in section 5.2, although the PC lattice is a bending dominated lattice, the struts of the PC lattice only sustain axial force when normal stress is applied to the lattice cell. Therefore, elastic buckling could happen to thin struts of the PC lattice before the plastic yield happens when the compressive axial force applied to the strut exceeds the critical elastic buckling force. Thus, a buckling constraint is developed to prevent the struts of the PC lattice cells from elastic buckling. In previous research, the Euler buckling formula, expressed as Eq. (29), is adopted to calculate the critical elastic buckling stress of an individual strut [9, 24].

$$\sigma_{cr} = \frac{\gamma \pi^2 E}{(l/k)^2} \quad (29)$$

Where the constant $\gamma$ depends on the boundary conditions of the individual strut. Because the boundary conditions of the lattice struts can be simplified as pin-jointed at the top and at the root of the strut, $\gamma = 1$ in Eq. (29). $l$ and $k$ are the length of the strut and the radius of gyration of the strut cross section, respectively. $l/k$ is the slenderness ratio of the strut. This slenderness radio is used in classifying struts according to length categories [40]. The Euler curve is shown in Figure 13. It indicates that when the slenderness ratio, $l/k$, of the strut is less than a certain value $\lambda_P$, the strut can be treated as a pure compression member without considering elastic buckling. However, due to unavoidable manufacturing defects, such as crookedness or load eccentricities, numerous tests relating to columns with intermediate-slenderness-ratios have proven that when the slenderness ratio, $l/k$, is less than $\lambda_Q$, the Johnson formula curve is more accurate than the Euler function to describe the column buckling behaviour, as illustrated in Figure 13 [40]. The Johnson formula has been widely used in the fields of machine, automobile, aircraft, and structural-steel construction designs. Due to the manufacturing limits of AM, the lattice strut cannot be overly thin. In fact, the slenderness ratios of lattice struts are smaller than $\lambda_Q$ in most engineering applications. The Johnson formula can be expressed as Eq. (30) [40].

$$\sigma_{cr} = \sigma_y^* - \left(\frac{\sigma_y^*}{2\pi}\frac{l}{k}\right)^2 \frac{1}{CE}, \text{where } \frac{l}{k} < \left(\frac{2\gamma\pi^2 E}{\sigma_y^*}\right)^{1/2} \quad (30)$$

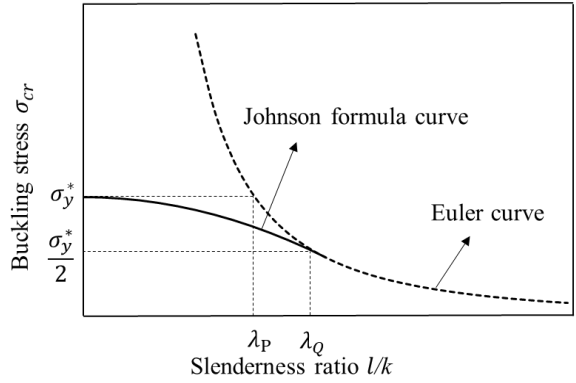

Figure 13.   Euler curve and Johnson formula curve for strut buckling

For a lattice strut with a circular section with radius $R$, the radius of gyration of the strut cross section $k$ is equal to $R/2$. Substituting $k$ in Eq. (29) and Eq. (30), noting the relative radius of strut $\bar{R} = R/L$, the buckling criterion for a lattice strut can be described as Eq. (31). The Euler formula is used to estimate the critical buckling stress of the lattice strut when its relative radius $\bar{R}$ is smaller than $\left(2\sigma_y^*/\gamma\pi^2 E\right)^{1/2}$. Otherwise, the Johnson formula is used for the lattice strut buckling analysis. For the PC lattice cell, since the struts are perpendicular to each other, the critical effective buckling compressive stress, $\sigma_{cr}^H(\bar{R})$, of the lattice cell can be calculated using Eq. (32), where S is the strut cross area and L is the length of the lattice cell. In optimization, the effective compressive stress of the lattice cell $\sigma_{ii}^H$ is constrained below $\sigma_{cr}^H$ to prevent strut buckling, as presented in Eq. (33), where $\sigma_{ii}^H$ is the effective compressive stress in direction-$i$, and $i$ = 1, 2, 3 for three-dimensional problems.

$$\sigma_{cr}^{strut} = \begin{cases} \frac{\gamma\pi^2 E}{4}\bar{R}^2, & \bar{R} < \left(\frac{2\sigma_y^*}{\gamma\pi^2 E}\right)^{1/2} \\ \sigma_y^* - \frac{\sigma_y^{*2}}{\gamma\pi^2 E \bar{R}^2}, & \bar{R} \geq \left(\frac{2\sigma_y^*}{\gamma\pi^2 E}\right)^{1/2} \end{cases} \quad (31)$$

$$\sigma_{cr}^H(\bar{R}) = \frac{S\sigma_{cr}^{strut}}{(L)^2} = \begin{cases} \frac{\gamma\pi^3 E}{4}\bar{R}^4, & \bar{R} < \left(\frac{2\sigma_y^*}{\gamma\pi^2 E}\right)^{1/2} \\ \pi\bar{R}^2\sigma_y^* - \frac{\sigma_y^{*2}}{\gamma\pi E}, & \bar{R} \geq \left(\frac{2\sigma_y^*}{\gamma\pi^2 E}\right)^{1/2} \end{cases} \quad (32)$$

$$\sigma_{ii}^H(\bar{R}, n) < \sigma_{cr}^H(\bar{R}) \quad (33)$$

## 6. Optimization problem standard formulation

The structural optimization problem in this study is defined as the problem of optimizing the distribution of lattice strut radii and fillet parameters to minimize structural compliance whilst satisfying the maximum weight fraction constraint, the yield constraint for both BCC and PC lattices, and the elastic buckling constraint for the PC lattice. Accordingly, the optimization problem formulation can be written as:

$$find\ \bar{R}, n$$

$$\bar{R} = [\bar{R}_1, \bar{R}_2, \cdots, \bar{R}_e, \cdots, \bar{R}_N]$$

$$n = [n_1, n_2, \cdots, n_e, \cdots, n_N]$$

$$\min J(\bar{R}, n) = \frac{1}{2} U^T K(\bar{R}, n) U \tag{34}$$

s.t

$$K(\bar{R}, n)U = F \tag{35}$$

$$C^H = C^H(\bar{R}_e, n_e) \tag{36}$$

$$\mathbf{K}(\bar{R}, n) = \sum_{e=1}^{N} \mathbf{\Gamma}_e^T \left( \int_{v_e} \mathbf{B}_e^T \mathbf{C}_e^H(\bar{R}_e, n_e) \mathbf{B}_e \, dv_e \right) \mathbf{\Gamma}_e \tag{37}$$

$$\frac{1}{N} \sum_{e=1}^{N} \bar{\rho}(\bar{R}_e, n_e) - V_f \leq 0 \tag{38}$$

$$\sigma^{Hill}(\bar{R}_e, n_e) - 1 \leq 0 \tag{39}$$

$$\frac{\sigma_{ii}^{cp}(\bar{R}_e, n_e)}{\sigma_{cr}^{H}(\bar{R}_e)} - 1 < 0, i = 1, 2, 3 \tag{40}$$

$$\bar{R}_{min} \leq \bar{R} \leq \bar{R}_{max} \tag{41}$$

$$0 \leq n \leq 5 \tag{42}$$

Where the vectors of the relative strut radii $\bar{R}$ and the fillet parameters $n$ are the design variables. $\bar{R}$ and $n$ are bounded between $\bar{R}_{min}$ and $\bar{R}_{max}$ and between 0 and 5, respectively. $N$ is the total number of lattice cells, and the subscript '$e$' denotes the $e$-th lattice cell in the design domain. The objective function is minimization of the total structural compliance $J(\bar{R}, n)$. $K(\bar{R}, \mathbf{n})$ and $U$ are the global stiffness matrix and the displacement vector, respectively. $F$ is the external load vector. $C^H(\bar{R}_e, n_e)$ represents the effective elastic tensor, developed in Section 4.3, for the $e$-th lattice cell. $\mathbf{B}$ is the element strain-displacement tensor. $\mathbf{\Gamma}_e$ is the connection matrix mapping the element stiffness matrix to the global stiffness matrix. $v_e$ is the volume of the $e$-th element. $V_f$ is the defined maximum volume fraction of the designed lattice structure. $\bar{\rho}(\bar{R}_e, n_e)$ denotes the relative density of the $e$-th lattice cell. $\sigma^{Hill}(\bar{R}_e, n_e)$ denotes the modified Hill's stress of the $e$-th lattice cell. $\sigma_{ii}^{cp}(\bar{R}_e, n_e)$ is the effective compressive stress of the $e$-th lattice cell in direction-$i$. For a three-dimensional optimization problem, $i = 1, 2, 3$. $\sigma_{cr}^{H}(\bar{R}_e)$ represents the critical elastic buckling stress constraint for the e-th lattice cell.

## 7. Design optimization framework for graded filleted lattice structures

An optimization framework is proposed to enable the optimal design of graded filleted lattice structures. A flowchart of the framework is shown in Figure 14. Firstly, samples of various combinations of lattice strut radii and fillet parameters are selected to create the database for building metamaterial models. Secondly, mesh sensitivity analysis is conducted to obtain the optimal mesh type and mesh size for numerical homogenization, taking both accuracy and efficiency into consideration. In the numerical homogenization step, FEA simulations are conducted to characterise the effective mechanical properties of the lattice RVEs, including the effective elastic tensor and modified Hill's yield stress. Thereafter, the polynomial regression is used to establish metamaterial models. The relative radii of lattice struts and fillet parameters are the independent design variables of the models. The effective lattice metamaterial properties, including relative density, effective elastic tensor and the modified Hill's yield stress, are all functions of the design variables. In addition, the elastic buckling constraint is developed for the PC lattice structures. The next steps are lattice structural optimization, post-processing, and detailed CAD model generation of the optimized graded filleted lattice structures. After optimization, the optimized strut radii and the optimized fillet parameters of the adjacent lattice cells could take different values. They can potentially result in sudden changes of strut radii at the joints of adjacent lattice cells, consequently causing stress concentration. To reduce the stress concentration, the optimal lattice strut radii and fillet parameters near the joints of the adjacent lattice cells are averaged in post-processing to ensure smooth geometric changes at the joints of adjacent lattice cells. The FEA simulations are conducted using ABAQUS (2018). The regression is conducted using MATLAB (2018). The optimization process is conducted using the commercial software Optistruct (2018). The method of feasible directions (MFD) algorithm is used as the optimization algorithm. The post-processing of the optimization results and the CAD model generation are conducted in MATLAB (2018) and Rhino (6.0), respectively.

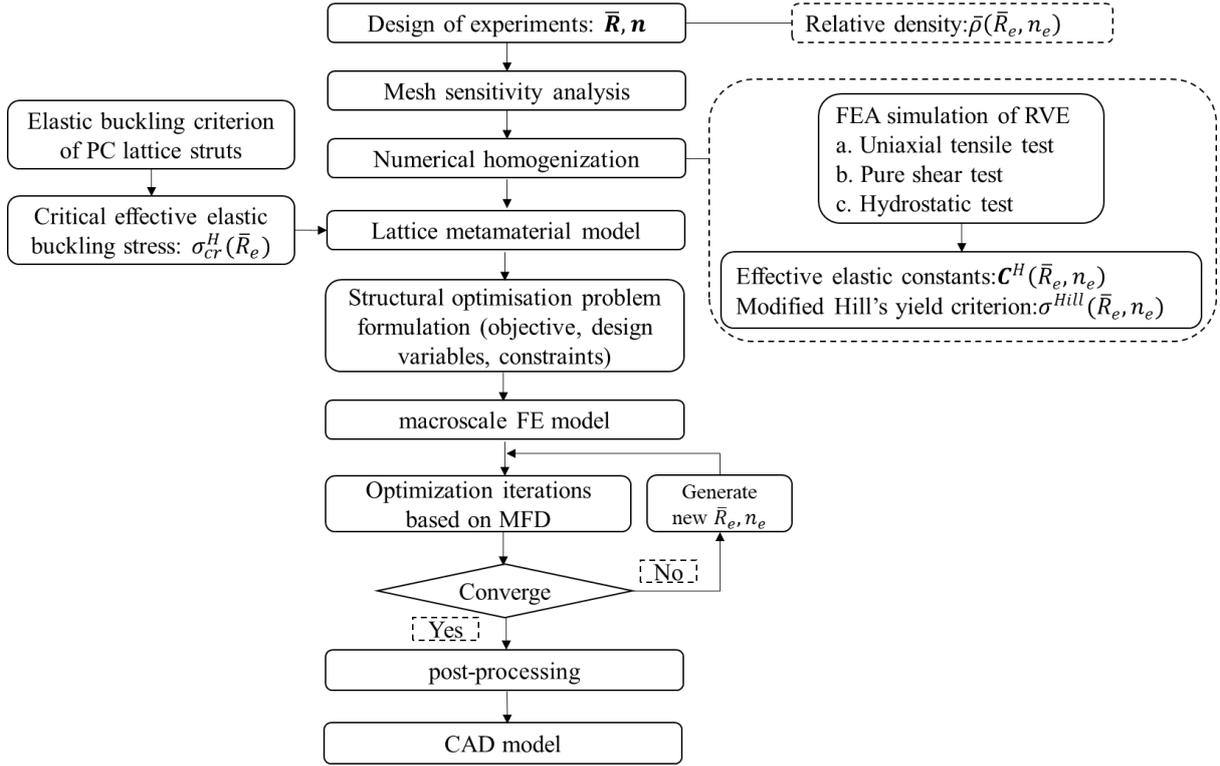

Figure 14. The flowchart of multiscale optimization of graded filleted lattice structure

## 8. Optimization case study

In this section, a Messerschmitt-Bolkow-Bolhm (MBB) beam case was designed to validate the graded filleted lattice structural optimization framework. Figure 15 illustrates the dimensions and boundary conditions of the beam. A force, with a magnitude of 24.3 KN along the direction-z, was uniformly applied to the central line (parallel to the y axis) of the upper surface of the beam. Two displacement constraints along direction-z were applied to the bottom surface of the beam with a distance of 20 mm from the two edges of the beam. The structural optimization problem is formulated following Eq. (34-42). The maximum volume fraction constraint $V_f$ is set to be 0.3. Based on the investigation in section 4.5, the maximum size of a lattice cell is defined as 5mm to ensure there are at least 8 lattice cells in one direction. The FE model of the MBB beam is shown as Figure 15. As mentioned earlier, Al 2024-T3 was used as the parent material to compose the lattice metamaterials.

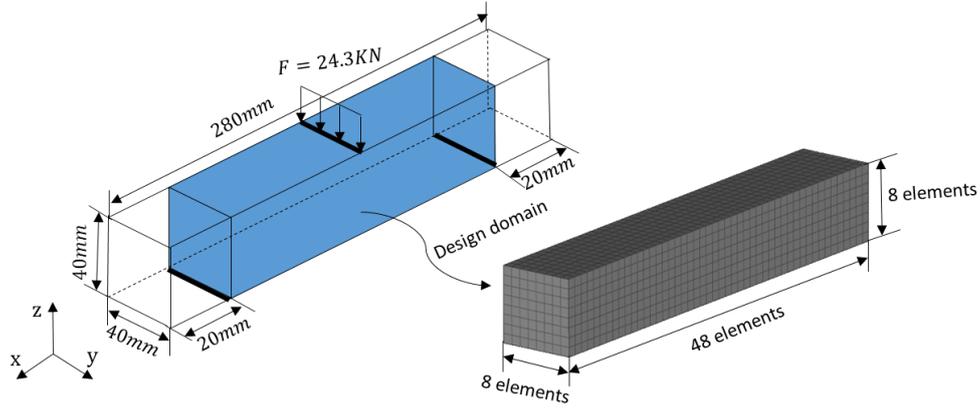

Figure 15. Dimensions and boundary conditions of the MBB beam optimization case

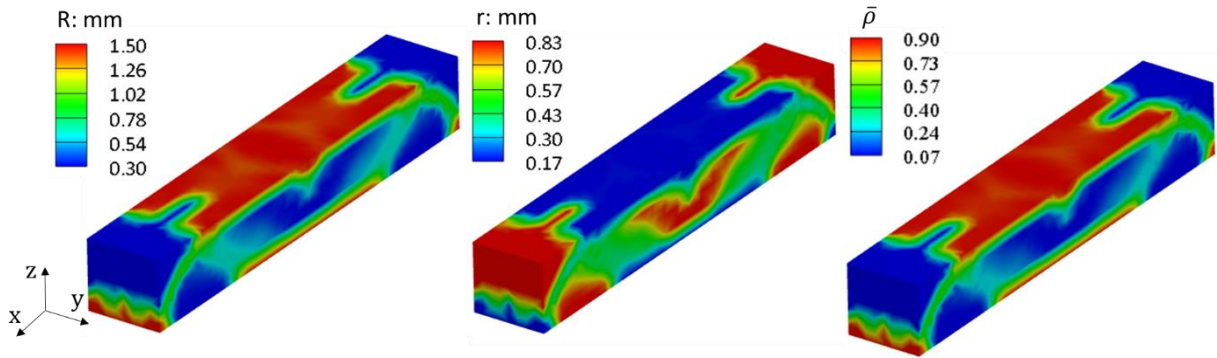

(a) MBB beam composed of BCC lattices

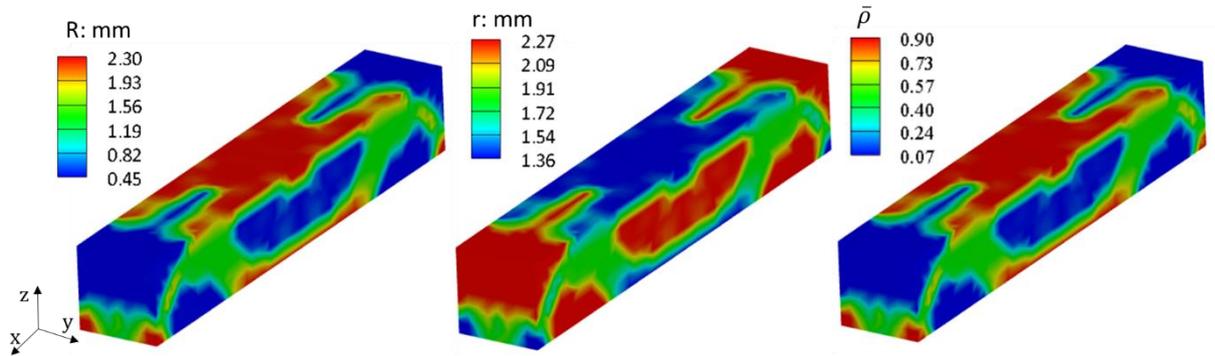

(b) MBB beam composed of PC lattices

Figure 16. The optimized distributions of strut radii, fillet radii and relative densities in the MBB beam composed of (a) BCC lattices, and (b) PC lattices.

Although the relative strut radii $\bar{R}$ and the fillet parameters $n$ are used in the optimization problem formulation for convenience, the actual values of the lattice strut radii $R$ and the fillet radii $r$ are used for CAD modelling. Therefore, Figure 16 shows the optimized distributions of the lattice strut radii, fillet radii, and relative densities in the MBB beams composed of the (a) BCC and (b) PC lattices. In general, the optimized relative densities of both the MBB beams show reasonable distributions, similar to an arch bridge shape. More materials are distributed at the top and bottom sections, while most materials are removed at the central section of the beam. Intermediate relative density regions are distributed along the loading paths between the top and bottom surfaces. For the beam composed of BCC lattices, there are more high relative density regions distributed at the top section; while for PC lattices, there are more materials with intermediate relative density connecting the top and the bottom surface of the beam. This difference could be caused by the difference between the effective elastic moduli of the PC and BCC lattices under a certain loading condition. For instance, the top and bottom sections respectively sustain compression and tension forces, which requires high effective Young's modulus to resist normal deformation. At the same relative density level, the PC lattice has a higher effective Young's modulus but a lower effective shear modulus, compared with the BCC lattice. So, to obtain the same level of Young's modulus in these regions, the BCC lattice metamaterial requires a higher relative density than the PC lattice metamaterial. For each beam, the distribution of lattice strut radii shows a very similar trend as the distribution of relative densities, because the design variable lattice strut radius is the main contributor to the relative density of a lattice cell; while the distribution of fillet radii shows an almost inversed trend compared with the distribution of relative densities. For instance, the value of $r$ is high where $\bar{\rho}$ is low. This is due to the fact that the effects of the design variable fillet parameter on improving the effective mechanical properties of both lattices are much more significant at a lower relative density, as illustrated in Figures 8 and 11.

Based on the optimized structures shown in Figure 16, Figure 17 (a) and (b) show the distributions of the modified Hill's stresses $\sigma^{Hill}(\bar{R}_e, n_e)$ corresponding to every lattice cell of the graded filleted BCC and PC lattice structures, respectively. The constraint Eq. (39) ensures that the yield criterion is satisfied, thus no modified Hill's stresses at any cell exceeds the value of 1. At the same time, as can be seen, there are a number of cells that have achieved $\sigma^{Hill}(\bar{R}_e, n_e) = 1$ for both structures, which means the yield constraint is active for both optimisation cases. In other words, the optimized structures are likely to yield under the given loading condition if the yield constraint was not defined. For engineering applications, a safety factor can also be integrated into Eq. (39) to constrain the $\sigma^{Hill}(\bar{R}_e, n_e)$ below a value smaller than 1.

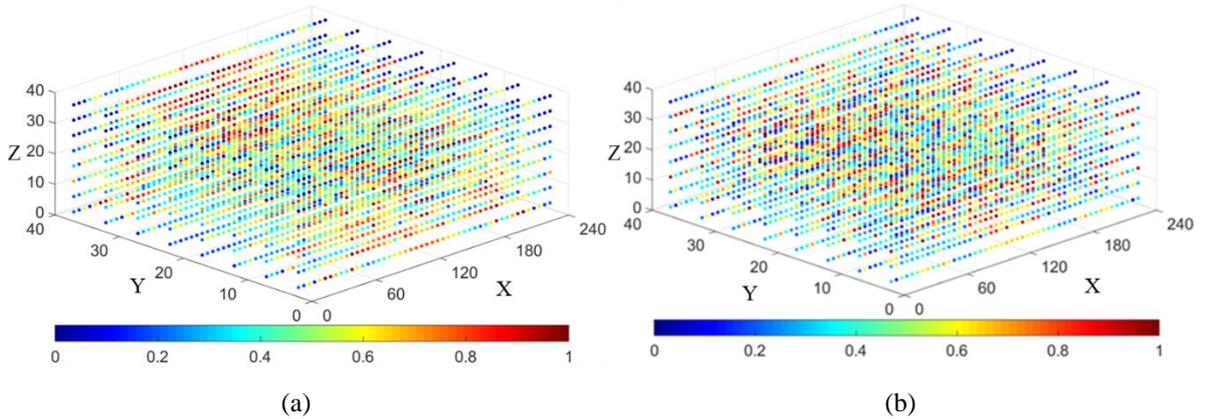

Figure 17.  Yield constraints of the optimized MBB (a) Modified Hill's stress distribution of BCC lattice structure. (b) Modified Hill's stress distribution of PC lattice structure.

To examine the buckling constraint for the optimisation of the PC lattice structure, a buckling factor is introduced:

$$Bf_e = \sigma_{ii}^H(\bar{R}_e, n_e)/\sigma_{cr}^H(\bar{R}_e), \qquad 0 \le n_e \le 5 \qquad (43)$$

According to Eq. (43), $Bf_e$ is constrained below 1. In this case study, since the stress in the direction-y is expected to be negligible, the buckling analysis is only conducted on the struts of the PC lattice structure along direction-x and direction-z. Based on the optimized structure shown in Figure 16, Figure 18 (a) and (b) illustrate the distributions of the buckling factors $Bf_e$ of the PC lattice strut along direction-x and direction-z, respectively, in the graded filleted PC lattice structure. For direction-x, high buckling factors are concentred at the top section of the beam, where the beam is mainly under compression along the direction-x. For direction-z, relatively high buckling factors are distributed around the middle section of the beam, where the vertical struts support the materials at the top and bottom regions to avoid beam collapse. The buckling constraint is not active in this case study but the buckling factors $Bf_e$ achieve up to 0.95, which still provides structural designers with insight on safety-critical regions. For engineering applications, a safety factor is suggested to be added to Eq. (40), in which case the buckling constraint is likely to become active.

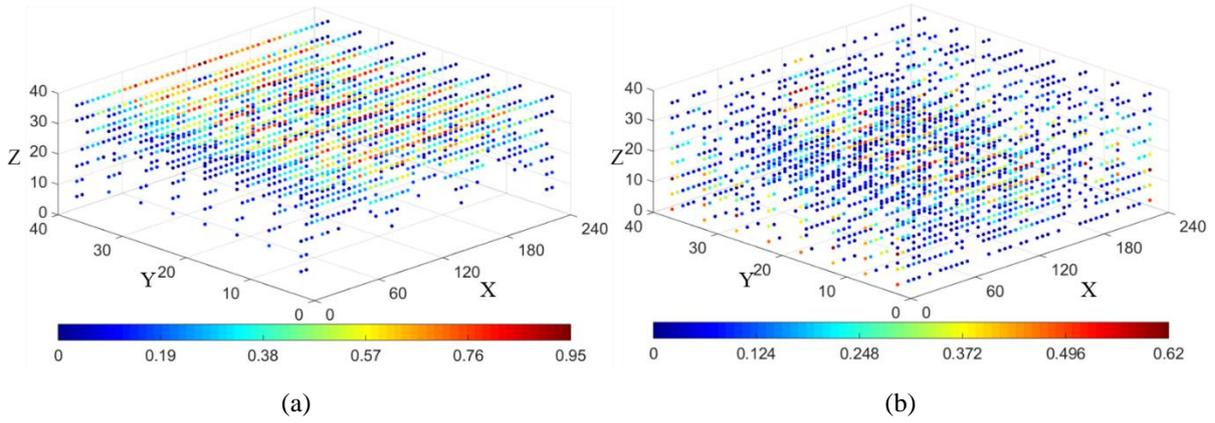

Figure 18. The distribution of the buckling factors ($Bf_e = \sigma_{ii}^H(\bar{R}_e, n_e)/\sigma_{cr}^H(\bar{R}_e)$) in the optimized MBB beam structure composed of PC lattices along (a) direction-x struts and (b) direction-z struts.

To demonstrate the effect of introducing fillets to the lattice joints on reducing the stress concentration in the optimized lattice structures, the optimization cases of the MBB beam composed of BCC lattices and PC lattices without filleted joints (i.e. $n = 0$) are conducted as comparisons. The case codes 'BCC' and 'PC' represent the non-filleted BCC and PC lattice structures, respectively. 'BCC-F' and 'PC-F' represent the filleted BCC and PC lattice structures, respectively. The optimized compliance for each case is normalized by the highest compliance value of all the cases. As discussed in Section 4.4, introducing filleted joints can improve the moduli of BCC and PC lattice metamaterials. Thus, it can be observed from Figure 19 (a) that the compliances of BCC-F and PC-F are reduced by 7% and 6% compared to BCC and PC, respectively, subject to the same loading condition and constraints. Figure 19 (b) illustrates the 100% stacked bar charts of the modified Hill's stresses of the four cases. Compared with the non-filleted lattice structures, the proportions of high modified Hill's stress region $\sigma^{Hill} \geq 0.95$ for both filleted lattice structures have declined. Additionally, the proportion of the modified Hill's stress region $0.85 \leq \sigma_e^{Hill} < 0.95$ in the BCC beam has also declined. This implies that the BCC lattices benefit more from the introduction of filleted joints compared with the PC lattices in the present case study.

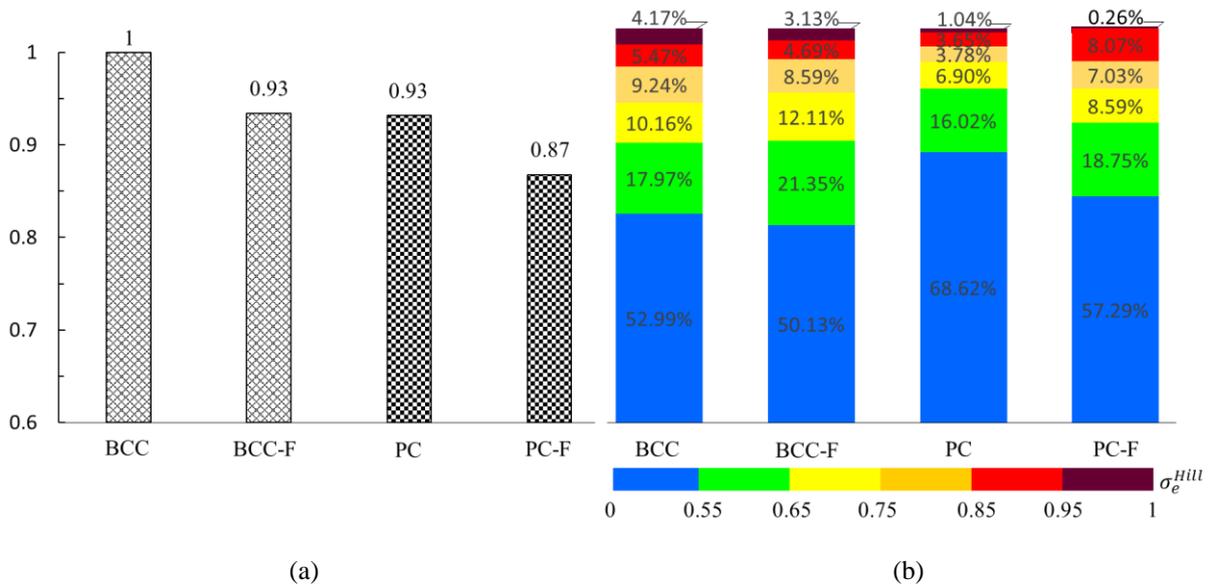

Figure 19. Optimized results comparison of the four cases (a) Comparison of normalized optimized compliance for different cases. (b) Comparison of Hill's stress interval ratio for different cases.

Finally, after post-processing of the optimized results, the CAD models of the optimized MBB beam composed of graded filleted BCC lattices and PC lattices are provided in Figure 20.

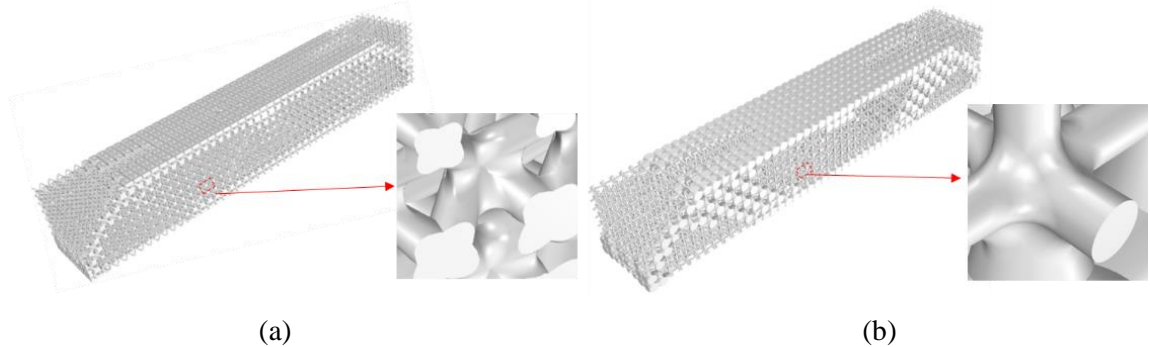

(a)                                                (b)

Figure 20. Construction of the CAD models of the optimized MBB beam composed of graded filleted (a) BCC lattices and (b) PC lattices.

## 9. Conclusion

In this work, an optimization framework is developed for the design and optimization of graded filleted lattice structures under the yield and strut elastic buckling constraints. This platform enables the simultaneous optimization of macroscale structures and the tailoring of lattice metamaterial properties satisfying the structural safety constraints.

The numerical homogenisation method is employed to efficiently obtain the effective mechanical properties of the defined quasi-isotropic lattice metamaterials, with different combinations of strut radii and fillet radii of both BCC and PC lattice RVEs. After FEA mesh and lattice cell sensitivity analyses, using ABAQUS, FEA simulations are conducted on a set of lattice RVEs, for each lattice type, as numerical experiments. The results of uniaxial tensile and pure shear testing are used to characterize the effective Young's modulus, effective shear modulus, and effective Poisson's ratios, so as to calculate the effective elastic constants; the results of uniaxial tensile, pure shear, and hydrostatic testing are used to characterize the effective uniaxial, shear, and hydrostatic yield stresses, respectively. Metamaterial models are developed, through polynomial regression using MATLAB, to quantify the relationships between each of the lattice metamaterial properties (i.e. relative density, effective elastic constants, and effective yield stresses) and the design variables (i.e. relative strut radii and fillet parameters). The response surfaces based on all determined models are smooth, with all R-squared values greater than 0.999. The analyses of the effects of filleted parameters show that introducing filleted joints can significantly improve the effective Young's modulus and uniaxial yield stress of the BCC lattices, and the effective shear modulus and shear yield stress of PC lattices, when the level of relative density is low (e.g. $\bar{\rho} = 0.2$).

A yield constraint, based on the modified Hill's yield criterion, is developed and determined as a function of the design variables (i.e. relative strut radii and fillet parameters), for the optimization of both the BCC and PC lattice structures. An elastic buckling constraint, based on the Euler buckling formula and the Johnson formula, is developed and determined as a function of relative strut radii, for the optimization of the PC lattice struts. Both yield and buckling constraints are integrated into an optimization problem formulation for the optimal design of graded filleted lattice structures.

An optimization framework is established to achieve optimized graded filleted lattice structures and generate CAD models, assisted by the commercial software Optistruct, MATLAB, and Rhino. A case study on minimising the compliance of an MBB beam composed of defined lattices is successfully carried out to demonstrate the optimization framework. The optimization results indicate that: (i) introducing fillets at joints can reduce the optimized structural compliance; (ii) introducing fillets at joints can reduce stress concentration and consequently reduce high modified Hill's stress regions in the optimized structures; and (iii) the yield and elastic buckling constraints guarantee the safety of the optimized structures, i.e. no yield or buckling occurred in the optimized beams.

Acknowledgement

The research was performed at the AVIC Centre for Structural Design and Manufacture at Imperial College London. Funding support from the Aviation Industry Corporation of China (AVIC), the First Aircraft Institute (FAI), and the China Scholarship Council (CSC) for this research is much appreciated.

**References**

[1] K. M. Abate, A. Nazir, Y.-P. Yeh, J.-E. Chen, and J.-Y. Jeng, "Design, optimization, and validation of mechanical properties of different cellular structures for biomedical application," *The International Journal of Advanced Manufacturing Technology,* vol. 106, no. 3-4, pp. 1253-1265, 2019.

[2] M. Helou and S. Kara, "Design analysis and manufacturing of lattice structures an overview," *International Journal of Computer Integrated Manufacturing,* vol. 31, no. 3, pp. 243-261, 2018.

[3] C. Imediegwu, R. Murphy, R. Hewson, and M. Santer, "Multiscale structural optimization towards three-dimensional printable structures," *Structural and Multidisciplinary Optimization,* 2019.

[4] R. Mines, "Parent materials and lattice characterisation for metallic microlattice structures," in *Metallic Microlattice Structures*: Springer, 2019, pp. 33-48.

[5] Dongseok *et al.*, "Multi-lattice inner structures for high-strength and light-weight in metal selective laser melting process - ScienceDirect," vol. 175, pp. 107786-107786.

[6] S.-I. Park and D. W. Rosen, "Homogenization of Mechanical Properties for Material Extrusion Periodic Lattice Structures Considering Joint Stiffening Effects," *Journal of Mechanical Design,* vol. 140, no. 11, 2018.

[7] L. Bai, C. Yi, X. Chen, Y. Sun, and J. Zhang, "Effective Design of the Graded Strut of BCC Lattice Structure for Improving Mechanical Properties," *Materials (Basel),* vol. 12, no. 13, Jul 8 2019.

[8] X. Cao, S. Duan, J. Liang, W. Wen, and D. Fang, "Mechanical properties of an improved 3D-printed rhombic dodecahedron stainless steel lattice structure of variable cross section," *International Journal of Mechanical Sciences,* vol. 145, pp. 53-63, 2018.

[9] J. Souza, A. Großmann, and C. Mittelstedt, "Micromechanical analysis of the effective properties of lattice structures in additive manufacturing," *Additive Manufacturing,* vol. 23, pp. 53-69, 2018.

[10] M. P. Bendsoe and N. Kikuchi, "Generating optimal topologies in structural design using a homogenization method," 1988.

[11] N. P. van Dijk, K. Maute, M. Langelaar, and F. van Keulen, "Level-set methods for structural topology optimization: a review," *Structural and Multidisciplinary Optimization,* vol. 48, no. 3, pp. 437-472, 2013.

[12] J. Liu *et al.*, "Current and future trends in topology optimization for additive manufacturing," *Structural and Multidisciplinary Optimization,* vol. 57, no. 6, pp. 2457-2483, 2018.

[13] J. Plocher and A. Panesar, "Review on design and structural optimisation in additive manufacturing: Towards next-generation lightweight structures," *Materials & Design,* vol. 183, 2019.

[14] S. Daynes, S. Feih, W. F. Lu, and J. Wei, "Optimisation of functionally graded lattice structures using isostatic lines," *Materials & Design,* vol. 127, pp. 215-223, 2017.

[15] A. Panesar, M. Abdi, D. Hickman, and I. Ashcroft, "Strategies for functionally graded lattice structures derived using topology optimisation for Additive Manufacturing," *Additive Manufacturing,* vol. 19, pp. 81-94, 2018.

[16] L. Cheng, J. Bai, and A. C. To, "Functionally graded lattice structure topology optimization for the design of additive manufactured components with stress constraints," *Computer Methods in Applied Mechanics and Engineering,* vol. 344, pp. 334-359, 2019.

[17]   Y. Wang, L. Zhang, S. Daynes, H. Zhang, S. Feih, and M. Y. Wang, "Design of graded lattice structure with optimized mesostructures for additive manufacturing," *Materials & Design,* vol. 142, pp. 114-123, 2018.

[18]   D. Li, W. Liao, N. Dai, and Y. M. Xie, "Anisotropic design and optimization of conformal gradient lattice structures," *Computer-Aided Design,* vol. 119, 2020.

[19]   Y. Wang, S. Arabnejad, M. Tanzer, and D. Pasini, "Hip Implant Design With Three-Dimensional Porous Architecture of Optimized Graded Density," *Journal of Mechanical Design,* vol. 140, no. 11, 2018.

[20]   E. Masoumi Khalil Abad, S. Arabnejad Khanoki, and D. Pasini, "Fatigue design of lattice materials via computational mechanics: Application to lattices with smooth transitions in cell geometry," *International Journal of Fatigue,* vol. 47, pp. 126-136, 2013.

[21]   Y. Tang, Y. Xiong, S.-i. Park, G. N. Boddeti, and D. Rosen, "Generation of Lattice Structures with Convolution Surface," presented at the Proceedings of CAD'19, 2019.

[22]   M. Dallago, S. Raghavendra, V. Fontanari, and M. Benedetti, "Stress concentration factors for planar square cell lattices with filleted junctions," *Material Design & Processing Communications,* vol. 2, no. 2, 2019.

[23]   G. N. Labeas and M. M. J. S. Sunaric, "Investigation on the Static Response and Failure Process of Metallic Open Lattice Cellular Structures," vol. 46, no. 2, pp. 195-204, 2010.

[24]   V. S. Deshpande, N. A. Fleck, and M. F. Ashby, "Effective properties of the octet-truss lattice material," *Journal of the Mechanics and Physics of Solids,* vol. 49, no. 8, pp. 1747-1769, 2001.

[25]   B. Ji, H. Han, R. Lin, and H. Li, "Failure modes of lattice sandwich plate by additive-manufacturing and its imperfection sensitivity," *Acta Mechanica Sinica,* vol. 36, no. 2, pp. 430-447, 2019.

[26]   S. L. Omairey, P. D. Dunning, and S. Sriramula, "Development of an ABAQUS plugin tool for periodic RVE homogenisation," *Engineering with Computers,* vol. 35, no. 2, pp. 567-577, 2018.

[27]   M. F. Ashby, "The properties of foams and lattices," *Philos Trans A Math Phys Eng Sci,* vol. 364, no. 1838, pp. 15-30, Jan 15 2006.

[28]   V. Deshpande, M. Ashby, and N. Fleck, "Foam topology: bending versus stretching dominated architectures," *Acta materialia,* vol. 49, no. 6, pp. 1035-1040, 2001.

[29]   J. Rossignac and A. Requicha, "Constant-radius blending in solid modelling," 1984.

[30]   M. A. Sanglikar, P. Koparkar, and V. Joshi, "Modelling rolling ball blends for computer aided geometric design," *Computer Aided Geometric Design,* vol. 7, no. 5, pp. 399-414, 1990.

[31]   T. Hermann, "Rolling ball blends and self-intersection," in *Curves and Surfaces in Computer Vision and Graphics III*, 1992, vol. 1830, pp. 204-209: International Society for Optics and Photonics.

[32]   O. Rehme, *Cellular design for laser freeform fabrication*. Cuvillier Verlag, 2010.

[33]   K. Ushijima, W. J. Cantwell, R. A. W. Mines, S. Tsopanos, and M. Smith, *An investigation into the compressive properties of stainless steel micro-lattice structures*. 2010.

[34]   R. Gümrük and R. A. W. Mines, "Compressive behaviour of stainless steel micro-lattice structures," *International Journal of Mechanical Sciences,* vol. 68, pp. 125-139, 2013.

[35]   G. Dong, Y. Tang, and Y. F. Zhao, "A 149 Line Homogenization Code for Three-Dimensional Cellular Materials Written in matlab," *Journal of Engineering Materials and Technology,* vol. 141, no. 1, 2018.

[36]   S. J. Hollister and N. Kikuchi, "A comparison of homogenization and standard mechanics analyses for periodic porous composites," *Computational Mechanics,* vol. 10, no. 2, pp. 73-95, 1992.

[37]   B. Hassani and E. Hinton, "A review of homogenization and topology optimization I—homogenization theory for media with periodic structure," *Computers & Structures,* vol. 69, no. 6, pp. 707-717, 1998.

[38] P. M. Suquet, "Elements of homogenization for inelastic solid mechanics, Homogenization Techniques for Composite Media," *Lecture notes in physics,* vol. 272, p. 193, 1985.

[39] Z. Xia, C. Zhou, Q. Yong, and X. Wang, "On selection of repeated unit cell model and application of unified periodic boundary conditions in micro-mechanical analysis of composites," *International Journal of Solids and Structures,* vol. 43, no. 2, pp. 266-278, 2006.

[40] R. G. Budynas and J. K. Nisbett, *Shigley's mechanical engineering design*. McGraw-Hill New York, 2008.